\documentstyle[eqsecnum,aps]{revtex}         % For Galley Format
\begin{document}

%\draft

\title{Dimensional Transmutation and
 Dimensional Regularization
\\
 in Quantum Mechanics
\\
{\small\bf I. General Theory}}

\author{Horacio E. Camblong}
\address{Department of Physics, University of San Francisco, San
Francisco, California 94117-1080
\\
\medskip
{\rm and}
}

\author{Luis N. Epele, Huner Fanchiotti,
 and Carlos A. Garc\'{\i}a Canal.}

\address{Laboratorio de F\'{\i}sica Te\'{o}rica,
 Departmento de F\'{\i}sica,
Universidad Nacional de La Plata,
\\ C.C. 67, 1900 La Plata, Argentina
\\
\bigskip
{\sc Dedicated to C.G. Bollini and J.J. Giambiagi} }

\def\diagram#1{{\normallineskip=8pt
       \normalbaselineskip=0pt \matrix{#1}}}

\def\diagramrightarrow#1#2{\smash{\mathop{\hbox to 
.8in{\rightarrowfill}}
        \limits^{\scriptstyle #1}_{\scriptstyle #2}}}

\def\diagramleftarrow#1#2{\smash{\mathop{\hbox to .8in{\leftarrowfill}}
        \limits^{\scriptstyle #1}_{\scriptstyle #2}}}

\def\diagramdownarrow#1#2{\llap{$\scriptstyle #1$}\left\downarrow
    \vcenter to .6in{}\right.\rlap{$\scriptstyle #2$}}

\def\diagramuparrow#1#2{\llap{$\scriptstyle #1$}\left\uparrow
    \vcenter to .6in{}\right.\rlap{$\scriptstyle #2$}}

\maketitle

\bigskip

\bigskip

%\begin{abstract}
\begin{quotation}
{\small
This is the first in a series of papers addressing the phenomenon
of dimensional transmutation in nonrelativistic quantum mechanics
within the framework of dimensional regularization.
Scale-invariant potentials are identified and their general
properties are derived. A strategy for dimensional renormalization
of these systems in the strong-coupling regime is presented, and
the emergence of an energy scale is shown, both for the
bound-state and scattering sectors. Finally, dimensional
transmutation is explicitly illustrated for the two-dimensional
delta-function potential.
}
%\end{abstract}
\end{quotation}

%\pacs{PACS numbers:}

%\narrowtext

%\newpage

\section{INTRODUCTION}
\label{sec:intro}

It is well known
 that, for various models of quantum field theory,
a mass scale emerges spontaneously through
 the renormalization procedure, even when the original
theory has no dimensional parameters.
This phenomenon, called dimensional transmutation, was first analyzed
in the 1973 seminal work of Coleman and Weinberg~\cite{col:73},
where the scalar field of massless scalar electrodynamics
was shown to develop a nonzero but arbitrary expectation value;
as a consequence, the particles of the theory
acquire nonzero physical masses~\cite{hua:82,kak:93}.
In short, the Coleman-Weinberg mechanism induces
radiative corrections to the Higgs potential,
thereby suggesting the relevance of dimensional transmutation
for the generation of particle masses~\cite{kak:93,wei:73}.

The main goal of this paper is to present a thorough investigation
of dimensional transmutation in nonrelativistic quantum mechanics,
with a threefold purpose in mind: (i) at the conceptual level, to
show that quantum field theory is not a prerequisite for its
existence; (ii) mathematically, to characterize the class of
scale-invariant potentials, as well as the subclass of potentials
that display dimensional transmutation; and (iii) at the practical
level, to develop useful tools for the treatment of a certain
class of singular quantum-mechanical potentials.

In particular, our work offers additional insight into two
problems that have been extensively studied in the literature: the
two-dimensional delta-function
potential~\cite{hua:82,tho:79,jac:91} and the inverse square
potential~\cite{mot:49,cas:50,lan:77,mor:53,jac:72,gup:93}.
Parenthetically, the family of delta-function potentials that is
discussed in this article is actually included in a larger class
of singular potentials whose apparent phenomenological usefulness
has been recognized for a long time, since the introduction of
pseudopotentials~\cite{wod:91} in the  early days of quantum
mechanics~\cite{bet:35} and  with subsequent applications of the
zero-range potential in nuclear physics~\cite{bre:47}, condensed
matter physics~\cite{kos:54}, statistical mechanics~\cite{lie:63},
 atomic physics~\cite{dem:89}, and particle physics~\cite{tho:79}.
Likewise, the inverse square potential
is related to the dipole potential, which has found applications in
molecular physics~\cite{lev:67}.
Even though earlier research on the subject had relied solely on
traditional quantum-mechanical techniques,
this situation has changed in recent years, with
the introduction of  numerous applications
of quantum field-theoretic tools and renormalization theory
to the same problems.
Specifically, among the many applications related
directly or indirectly to our singular potentials,
the following are worth mentioning:
(i) the mathematical
formulation of the theory of pseudopotentials~\cite{alb:88},
which started with  the works
of Ref.~\cite{zel:60} and
includes the technique of self-adjoint
extensions~\cite{jac:91,gol:77,ger:89a};
(ii) the study of the nonrelativistic
limit of the $\phi^{4}$ theory, with the concomitant
question of its triviality~\cite{beg:85};
(iii) the basic conceptual understanding of quantum field theory
mechanisms in the simpler framework of
quantum mechanics,
including standard regularization and
renormalization of the singular potentials mentioned
above~\cite{jac:91,gup:93,gos:91,hol:93,mea:91,adh:95a,cab:96,phi:98a},
anomalies~\cite{jac:91,hol:93,cab:96},
renormalization group
analysis~\cite{gup:93,gos:91,adh:95a,boy:94,fie:96,pol:96},
and effective field theory approach~\cite{phi:98a,lep:97,per:99};
(iv) the analysis of (2+1)-dimensional theories, including
gravity~\cite{ger:89b}, as well as
Chern-Simons theory~\cite{jac:90}, the Aharonov-Bohm
effect~\cite{ger:89a,hag:90a}, and
the dynamics of
anyons~\cite{man:91};
(v) new applications of contact potentials in condensed matter
physics, e.g., for the quantum Hall effect~\cite{cav:98};
and
(vi) the modern formulation, using effective field
theory~\cite{wei:79},
of the nucleon-nucleon
potential~\cite{wei:90}, which has led to a
plethora of quantum-mechanical
pseudopotentials~\cite{phi:98a,kap:97}.
Our paper  naturally follows the
trend set by this extensive bibliography.

The main focus of our work will be on the concept of {\em
dimension\/}, a term that has been extensively used in the physics
literature to describe two conceptually distinct ideas. The
meaning to which dimensional transmutation refers is that which
relates to measurement and which characterizes the class of
physical quantities that exhibit a certain type of power-law
behavior (dimensional homogeneity) with respect to a given choice
of fundamental quantities~\cite{buc:14}. In what follows, we will
use the term dimensionality to denote the exponents of the
associated homogeneous behavior for any given physical
quantity~\cite{wei:95}. At first sight, dimensional transmutation
is paradoxical, because it produces a scale in a problem devoid of
dimensional parameters, in apparent violation of Buckingham's
$\Pi$ theorem of orthodox dimensional analysis~\cite{buc:14}.
However, this paradox can be ultimately resolved by invoking the
dimensional arbitrariness intrinsic in the renormalization
framework~\cite{ste:81}.

The other widely used acceptation of dimension refers to a
 geometric concept, a property of the space where events take place
(e.g., space-time in
relativistic physics).
In this paper we will extensively exploit the trivial connection between
these two concepts, that is,
that the dimensionality of an element of volume in a given space,
expressed in terms of units of length,
is equal to its geometric dimension.
This connection has been largely used in
the renormalization of relativistic quantum field theories,
where it is further reinforced by
the implementation of dimensional
regularization~\cite{bol:72,lei:75}.
Correspondingly,
we will use dimensional regularization as a natural technique
that renders obvious the
spontaneous generation of a dimensional scale
for a scale-invariant theory.

Our paper is organized as follows. In Section~\ref{sec:DA} we discuss
the meaning of dimensional transmutation in terms of dimensional
analysis and renormalization theory.
Section~\ref{sec:scale_invariant} is devoted to the concept of
scale-invariant potentials in nonrelativistic quantum mechanics,
where generic properties related to scaling and scale symmetry are
derived. Section~\ref{sec:DT_via_DR} establishes a general
framework for the regularization of scale-invariant potentials
based upon dimensional continuation; this procedure is later
extended to a renormalization scheme in
Section~\ref{sec:renormalization}, where dimensional transmutation is
shown to arise in the strong-coupling regime. Finally, an
application of the theory is illustrated in Section~\ref{sec:delta},
which deals with some aspects of the two-dimensional
delta-function potential. The conclusions of our analysis are
presented in Section~\ref{sec:conclusions}. The appendices deal with
the necessary results regarding $D$-dimensional Euclidean spaces,
Green's functions, and scattering.

Additional applications for rotationally invariant problems
will follow in the second paper in this series~\cite{cam:00b}.

\section{DIMENSIONAL ANALYSIS, RENORMALIZATION, AND DIMENSIONAL
TRANSMUTATION}
\label{sec:DA}

The description of a physical system in the context
of a given theory, either in terms of a Lagrangian or of a
Hamiltonian, includes a certain number of parameters.
Usually, their values may all be fixed from the start by
the laws of nature, in the form of ``constants,''
but one always enjoys the mathematical freedom to make some  of
them become variable parameters as needed.
Then, for the discussion that follows,
these parameters will arbitrarily be classified into two groups:
``constant'' or fundamental and ``variable'' or dynamical.
By varying the variable parameters one introduces a whole
 class ${\mathcal C}$ of physical systems,
all characterized by the same values of the
``constants''~\cite{parameters}.

Fundamental parameters are those that are fixed constants  for
all the members of the given class of systems.
Needless to say, they are dimensional because ``fundamental''
dimensionless parameters amount to
plain numerical constants.
Typical fundamental parameters of choice are the dimensional universal
constants of nature, for example, $\hbar$ and $c$.

The second group is composed of the dynamical parameters
that acquire different values
for the different members of the class ${\mathcal C}$.
As an example we could mention the masses of  particles
or the coupling constants of the interactions.

The reduction in the number of dimensionally independent
quantities can be accomplished by arbitrarily assigning particular
numerical values to a subset of the fundamental parameters. This
procedure amounts to the selection of a generalized {\em natural
system of units\/}, in which the number of fundamental dimensions
is reduced. For example, in relativistic quantum field theory, it
is customary to choose $c=1$ and $\hbar=1$, so that the theory is
described in terms of a single fundamental dimension---usually
taken as inverse length $\Lambda=L^{-1}$, which is equivalent to
mass, momentum, and energy. Even though all the physical
dimensions can be restored easily at any stage of the calculation,
it is clear that great simplification is achieved in the
dimensional analysis of various physical quantities.

Similarly, in one-particle nonrelativistic quantum mechanics, one
has the freedom to use $\hbar$ and $m$ (where $m$ is the
particle's mass) as fundamental parameters that define a
particular generalized natural system of units; in this paper, we
will choose $\hbar=1$ and $m=1/2$. Then, we will be left with only
one fundamental dimension, which we will take again as inverse
length $\Lambda=L^{-1}$ or momentum. Consequently, in what
follows, we will define the inverse-length dimensionality $q={\rm
dim} \left[ Q \right]$ of a physical quantity $Q$ as the exponent
that expresses its physical dimension $\Lambda^{q}$ in terms of
inverse length, that is,
\begin{equation}
q= {\rm dim} \left[ Q  \right] = \frac{\Lambda}{ \left[ Q \right]
} \frac{ \partial \left[  Q \right] } {\partial \Lambda}  \;  .
\end{equation}
For nonrelativistic quantum mechanics,
Table~\ref{tab:dimensions}
summarizes the dimensionalities of the most common
physical quantities.

\newtheorem{tab}{\rm TABLE}
\renewcommand{\thetab}{\Roman{tab}}

\begin{center}
\begin{tab}
\hspace{-.05in}. \hspace{.1in}
{\rm Physical Dimensions of Various Physical Quantities.}
\label{tab:dimensions}
\end{tab}
\end{center}

\begin{center}
\begin{tabular}{cccc}  \hline\hline
{  Physical quantity} & { Ordinary dimensions} &
{ ``Natural'' dimensions} &
 { Dimensionality} \\  \hline
Length & $L$ & $L$ & -1 \\
Time & $T$ & $L^{2}$ & -2 \\
Velocity & $LT^{-1}$ & $L^{-1}$ & 1 \\
Linear momentum & $MLT^{-1}$ & $L^{-1}$ & 1 \\
Angular momentum & $ML^2T^{-1}$ & 1 & 0  \\
Energy  & $ML^2T^{-2}$ & $L^{-2}$ & 2  \\
Cross section & $L^{D-1}$  & $L^{D-1}$ & $-\left( D-1 \right)$  \\
Wave function (normalized) & $L^{-{D}/2}$
& $L^{-{D}/2}$ & $D/2$
\\
\hline\hline
\end{tabular}
\end{center}

{\small
{\em Note\/}.
``Natural'' dimensions are defined by the choice $\hbar=1$ and
$2m=1$. The  geometric dimension of
position space is $D$.}

\bigskip

Let us now explore the consequences of the possible existence of
dimensional parameters. For a given physical system, characterized
by a Lagrangian or a Hamiltonian, oftentimes there exists at least
one dimensional parameter, which can be used to define a
system-specific or intrinsic scale. To illustrate how this is
done, let us consider the nonrelativistic quantum-mechanical
dynamics of a single particle in one dimension, such that the
external-interaction potential contributes only one dimensional
parameter $\lambda$; for example, for an attractive power-law
potential, one may study the possible existence of bound states
through the  Schr\"{o}dinger equation
\begin{equation}
\left[ - \frac{d^{2}}{dx^{2}} + {\rm sgn} \left( \beta \right)
\lambda \, |x|^{\beta}  \right]
\,  \Psi (x)
=
E \,  \Psi (x)
\;  .
\label{eq:Schr_ex}
\end{equation}
Dimensional analysis shows that
 ${\rm dim} \left[ \lambda \right] = \ell =2+\beta$; then
$\lambda^{1/\ell}$ will define a basic unit of inverse
length or momentum.
Any dimensional quantity Q of dimensionality q will then be equal
to $\lambda^{q/\ell}$,
up to a numerical factor;
 similarly, a function $Q(x)$ of position
(or $Q(p)$ of momentum),  with dimensionality q, will then be equal
to $\lambda^{q/\ell}$ times a dimensionless function
of $\lambda^{1/\ell} x$ (or of $\lambda^{-1/\ell}  p$).
In particular, a characteristic ground-state energy may be estimated
as $\lambda^{2/(2+\beta)}$.
In other words,
dimensional analysis gives nontrivial information about the system.

The obvious statements of the previous paragraph
can be summarized in the
$\Pi$ theorem of dimensional analysis~\cite{buc:14},
which we state here without proof, with the
intention of generalizing it later in this section.
Consider a physical phenomenon described  by
$M$ dimensional characteristic
parameters $a_{1}, \ldots, a_{M}$,
such that $R$ of them are dimensionally independent.
Then, given an equation
\begin{equation}
F(a_{1}, \ldots, a_{M})=0
\label{eq:DA_variables}
\end{equation}
involving these $M$ parameters, there exist $N$ independent
dimensionless power products
$\Pi_{1}, \ldots, \Pi_{N}$ of $a_{1}, \ldots, a_{M}$,
such that Eq.~(\ref{eq:DA_variables}) is equivalent to
\begin{equation}
\Phi (\Pi_{1}, \ldots, \Pi_{N})=0
\;  ,
\label{eq:Pi}
\end{equation}
with
\begin{equation}
N= M-R
\; .
\label{eq:Pi_DOF}
\end{equation}

For example, working in a natural system of units with only one
independent dimension, it follows that
$N  = M-1$, which describes the situation
of the previous paragraph.

But what happens if the system exhibits no explicit dimensional
dynamical parameter at the level of the Lagrangian or Hamiltonian?
As we will see in Section~\ref{sec:scale_invariant}, such a system is
{\em scale-invariant\/}. An example is the power-law potential $-
\lambda |x|^{-2}$ ($\beta=-2)$ because its coupling $\lambda$ is
dimensionless. Then, naive dimensional analysis is at a loss to
make any meaningful predictions. In this case, if a new scale
arises (for example, a bound state under
 a scale-invariant potential),
dimensional analysis implies that it has the following
properties:
\begin{itemize}
\item
It is {\em spontaneously generated\/}, in the sense that it
characterizes the solution of a theory that is scale-invariant at
the level of the classical Lagrangian or Hamiltonian. This amounts
to an instance of quantum-mechanical breaking of classical scale
symmetry--also called {\em scale
anomaly\/}~\cite{jac:72,jac:90,tre:85}.
\item
It is totally {\em arbitrary\/}
because no privileged value is defined a priori within the theory.
If it were not arbitrary, it would violate the $\Pi$ theorem
 in an irreconcilable way.
\end{itemize}
This manifestation of an arbitrary
and spontaneously generated scale in a scale-invariant theory is
known as dimensional transmutation~\cite{col:73}.

In short, in the solution to a well-posed question within the
scale-invariant theory, a dimensionally transmuted scale $B$ may
appear spontaneously.
 How does it come into existence in apparent violation of
naive dimensional analysis? Our goal is to disentangle the
mechanism that leads to this transmutation. This will be
implemented by means of a regularization-renormalization
procedure. The regularization technique introduces a dimensional
parameter $\mu$, in terms of which the scale $B$ is expressed.
Thus, a dimensional transfer takes place, whereby a dimensionless
parameter $\lambda$ is ``transmuted'' into or traded for a
dimensional scale $B$. This simple process can be  represented
diagrammatically in the form
\begin{center}
\begin{tabular}{ccc}  
\hspace{.2in} {Initial problem} \hspace{.2in}
 &
\hspace{.2in}
{Technique}
\hspace{.2in} & {Solution}
\hspace{.2in}
\\ \hline
{\it Lagrangian/Hamiltonian\/}
& {\it Regularization/renormalization\/} &
 {\it Physical quantity\/} \\
 dimensionless  & arbitrary & measurable \\
coupling $\lambda$ &
$\mbox{ \hbox to 2in{\rightarrowfill} }
\atop \mbox{dimensional scale} \;  \;  \;  \mbox{\large $\mu$}$
& dimensional scale $B$ \\
& & \\
\end{tabular}
\end{center}

We conclude this section by stating the modification of
orthodox dimensional analysis needed to encompass this anomalous 
behavior.
As discussed in Ref.~\cite{ste:81},
the usual assumption underlying the $\Pi$ theorem
is that the function $F(a_{1},\ldots,a_{M})$
of Eq.~(\ref{eq:DA_variables}) is {\em uniquely\/}
defined, an assumption that breaks down
when the Lagrangian does not describe a single
theory but a class of theories parametrized with
renormalization parameters.
This manifests itself in a
 theory that is ill-defined or exhibits singularities
of some sort, in which case
the Lagrangian or Hamiltonian cannot represent a complete description
of the physics; thus, renormalization is needed.
 When the number of
independent sliding scales
or renormalization parameters
is $\sigma$, the required modification is
of the $\Pi$ theorem is obviously
\begin{equation}
N= M+\sigma-R
\; .
\label{eq:Pi_DOF_mod}
\end{equation}
Equation~(\ref{eq:Pi_DOF_mod}) states that the number of
``available'' variables is $M'=M+\sigma$, rather than $M$; in
particular, it provides the necessary freedom to permit the
emergence of dimensional transmutation. The framework for deriving
conclusions directly from Eq.~(\ref{eq:Pi_DOF_mod}) will be
referred to as generalized dimensional analysis.

\section{CHARACTERIZATION OF SCALE-INVARIANT POTENTIALS}
\label{sec:scale_invariant}

\subsection{Scale Symmetry and Homogeneity}
\label{sec:scale_symmetry}

In this section we set out to define and characterize
mathematically the class of scale-invariant potentials $V({\bf
r})$ in one-particle nonrelativistic quantum physics. In a strict
sense, we are referring to a physical system whose classical
action
 \begin{equation}
 S = \int \left[
 \frac{1}{2} \,
 m \stackrel{.}{\bf r}^{2}
  - V({\bf r}) \right] \,
 d t
 \;
\end{equation}
is invariant under the scale transformations ${\bf r} \rightarrow
\varrho {\bf r}$, $t \rightarrow \tau t$ (with $\varrho>0$ and
$\tau>0$). This scale symmetry is satisfied if and only if each
one of the two terms of the nonrelativistic action---the
kinetic-energy term $\int d t \, m \stackrel{.}{\bf r}^{2} \! /2$
and the potential-energy term $-\int d t \, V ({\bf r}) $---are
left unchanged. Due to the spatial and time dependence of the
nonrelativistic kinetic-energy term, this invariance condition is
satisfied only when $\varrho^{2}= \tau$ (obviously consistent with
the dimensional analysis of Table~\ref{tab:dimensions}), while the
invariance of the potential-energy term requires that
\begin{equation}
V(\varrho {\bf r})= \varrho^{-2} V({\bf r})
\;  .
\label{eq:homogpotential_action}
\end{equation}
As Eq.~(\ref{eq:homogpotential_action}) is valid for all
$\varrho>0$, the class of scale-invariant potentials is identical
to that of {\em homogeneous potentials of degree -2\/}. As we will
see next, this is the same condition to be satisfied when the
potential does not exhibit any explicit dimensional scale.

\subsection{Dimensional Scaling in Nonrelativistic Quantum Mechanics}
\label{sec:nrQM}

One-particle nonrelativistic quantum mechanics in the presence of
a stationary potential $V({\bf r})$ is described in the
$D$-dimensional position-space representation of the
Schr\"{o}dinger picture via the solutions of the time-independent
Schr\"{o}dinger equation, which in natural units reads
\begin{equation}
\left[ - \nabla^2 + V({\bf r}) \right]
\,  \Psi ({\bf r})
=
E \,  \Psi ({\bf r})
\;  .
\label{eq:Schr1}
\end{equation}

The analysis and interpretation of the solutions to
Eq.~(\ref{eq:Schr1}) will become more transparent when the
transition to its dimensionless version is carried out. This can
be accomplished by rescaling all quantities appearing in
Eq.~(\ref{eq:Schr1}) by means of a dimensional parameter $\mu$,
which we will assume to represent an inverse-length standard;
then, $\mu$ will satisfy the properties: (i) inverse-length
dimensionality,
\begin{equation}
{\rm dim} \left[ \mu \right]=1
\; ;
\end{equation}
(ii)
positivity,
\begin{equation}
\mu>0
\; .
\label{eq:positivity}
\end{equation}

In general, there are many possible characteristic scales that may
serve as $\mu$: they could either be intrinsic to the system or
arbitrary scales introduced via regularization.
In any case, we will not be concerned with the multi-scale case,
because
our ultimate goal is to analyze the extreme scenario where there is no
intrinsic dimensional parameter, but an arbitrary sliding scale
$\mu$ is introduced by the regularization
procedure.
Then any physical quantity $Q$ of dimensionality $q$
will be equal to a numerical coefficient times $\mu^q$, whence
its dimensionless counterpart will be defined as $\mu^{-q} Q$.
If, in addition, the quantity is a function of either
position or momentum, it will be of the form
$\mu^q$ times a dimensionless function of the dimensionless
position
\begin{equation}
 \mbox{\boldmath $ \xi$}
                            =\mu \, {\bf r}
\; ,
\label{eq:dimensionlessr}
\end{equation}
 or of the dimensionless momentum
\begin{equation}
 \mbox{\boldmath $ \pi$}
                            =\mu^{-1} \, {\bf p}
\;   .
\label{eq:dimensionlessp}
\end{equation}
Correspondingly, dimensional analysis predicts that the potential
energy function (whose dimensionality is 2) should
have a dependence on the dimensional parameter $\mu$ given by
\begin{equation}
V({\bf r}, \mu)= \mu^2 \,
{\mathcal V}
 (\mu {\bf r})
\;  ,
\label{eq:dimensionlessV1}
\end{equation}
for arbitrary $\mu$.
In particular, one can define the reduced function
${\mathcal V} ( \mbox{\boldmath $ \xi$} )$, via
Eq.~(\ref{eq:dimensionlessV1}), in the form
\begin{equation}
{\mathcal V} ( \mbox{\boldmath $ \xi$} )
=
\mu^{-2} \, V
( \mu^{-1} \mbox{\boldmath $ \xi$} , \mu)
\; ,
\label{eq:dimensionlessV2}
\end{equation}
or straightforwardly
by specializing
 to the unit value of the dimensional parameter, i.e.,
\begin{equation}
{\mathcal V}
( \mbox{\boldmath $ \xi$} )
=
V
( \mbox{\boldmath $ \xi$} , \mu=1)
\; .
\label{eq:dimensionlessV3}
\end{equation}
The right-hand side of
Eq.~(\ref{eq:dimensionlessV1}) displays the
two sources of possible scale dependence of the potential:
the one associated with the dimensionality of $V$ as a
potential energy (i.e., $\mu^2$)
and the one associated with the functional form of the potential
(as described by ${\mathcal V} (\mu {\bf r})$).
Equation~(\ref{eq:dimensionlessV1}) implies
that
\begin{equation}
\mu \frac{\partial \, |V({\bf r},\mu)| }{\partial \mu}
=
2 |V({\bf r},\mu)| + {\bf r} \cdot \mbox{\boldmath ${\nabla}$}
 |V({\bf r}, \mu) |
\;  , \label{eq:Euler1}
\end{equation}
where the first term on the right-hand side is the dimensionality of
 the potential energy
and the second term represents the degree of the functional
dependence of the potential energy with respect to the given scale.
For example, for a power-law potential
 $V({\bf r}) \propto  r^{\beta}$, the functional dependence amounts to
${\mathcal V}( \mbox{\boldmath $ \xi$}) \propto \xi^\beta$, whence
 $V({\bf r}, \mu) \propto  \mu^{2+\beta}$,
which describes the {\em total\/} scale dependence of the
potential
energy function under arbitrary rescaling.
Notice that, for $\beta=-2$,
 $V({\bf r}, \mu)$ is independent of $\mu$; i.e., it is
scale-independent (see the next section).
Table~\ref{tab:dimensionlessquant}
 gives a list of the various dimensionless quantities
of interest.
\begin{center}
\begin{tab}
\hspace{-.05in}. \hspace{.1in}
{\rm Dimensionless Counterparts  of Various Physical Quantities
in Nonrelativistic Quantum Mechanics.}
\label{tab:dimensionlessquant}
\end{tab}
\end{center}

\begin{center}
\begin{tabular}{cccc}  \hline\hline
{  Physical quantity} & { Symbol} &
 { Dimensionality} &
{ Dimensionless form
}  \\  \hline
Position  & ${\bf r}$ & -1 & $  \mbox{\boldmath $ \xi$}
                            =\mu \, {\bf r}$ \\
Linear momentum   & ${\bf p}$ & 1 & $  \mbox{\boldmath $ \pi$}
                            =\mu^{-1} \, {\bf p}$ \\
Kinetic energy & $- \nabla_{\bf r}^2 $ & 2 &
                   $- \nabla_{
\mbox{\boldmath $\scriptstyle{ \xi}$} }^2 =
                     -\mu^{-2}  \nabla_{\bf r}^2 $ \\
Potential energy & $V({\bf r}) $ & 2 & $ {\mathcal V}
( \mbox{\boldmath $ \xi$} ) $
=
  $\mu^{-2} V(\mu^{-1} \mbox{\boldmath $ \xi$} , \mu )
$ \\
Energy  & $E$ & 2 & $ \eta = \mu^{-2} E $\\
Wave function (normalized) & $\Psi ({\bf r})$
& $D/2$
  & $\Phi (  \mbox{\boldmath $ \xi$} )
        = \mu^{-{D}/2}
\Psi (\mu^{-1}  \mbox{\boldmath $ \xi$} ) $
\\
\hline\hline
\end{tabular}
\end{center}

Rescaling of Eq.~(\ref{eq:Schr1})
with the parameter $\mu$ yields
\begin{equation}
\left[
- \nabla_{\mbox{\boldmath $\scriptstyle{\xi}$}}^2 +
{\mathcal V}
( \mbox{\boldmath $ \xi$} )  \right]
\,  \Phi (  \mbox{\boldmath $ \xi$} )
= \eta
 \, \Phi (  \mbox{\boldmath $ \xi$} )
\;  ,
\label{eq:Schr2}
\end{equation}
which
describes an eigenvalue problem
for the dimensionless eigenfunctions $  \Phi (  \mbox{\boldmath $ \xi$} 
) $,
with dimensionless eigenvalues
\begin{equation}
\eta = \mu^{-2} E
\;   ,
\label{eq:dimensionlessE}
\end{equation}
in a space of arbitrary dimension $D$.
In addition, it is convenient to normalize the wave function
$\Phi (  \mbox{\boldmath $ \xi$} )$
with respect to its dimensionless argument
$\mbox{\boldmath $ \xi$} $, that is,
\begin{equation}
| \Phi (  \mbox{\boldmath $ \xi$} ) |^{2}
\, d^{{D}} \xi
        =  | \Psi (\mu^{-1}  \mbox{\boldmath $ \xi$} ) |^{2}
\, d^{{D}} r
\;  ,
\end{equation}
a condition that yields the rescaling
\begin{equation}
 \Phi (  \mbox{\boldmath $ \xi$} )
        = \mu^{-{D}/2}
\Psi (\mu^{-1}  \mbox{\boldmath $ \xi$} )
\;  .
\label{eq:dimensionlesspsi}
\end{equation}

As usual, the solutions
to Eq.~(\ref{eq:Schr2}) should be separately
obtained and interpreted for
the bound-state and scattering sectors.
Additional conclusions about
these specific problems will be drawn in
Subsections~\ref{sec:BS_sector_renormalized},
\ref{sec:scattering_sector_renormalized},
and \ref{sec:DA_renormalized}.

\subsection{Absence of Explicit Dimensional Scales and Homogeneity 
Revisited}
\label{sec:scale_invariant2}

Let us now characterize the class of potentials that do not
exhibit any dimensional scale, in a space of arbitrary dimension
$D$. We will resort to the general framework developed in
Subsection~\ref{sec:nrQM}, where we considered potentials that depend
upon only one dimensional parameter or none (once the choice
$\hbar=1$ and $2m=1$ has been made); from
Eq.~(\ref{eq:dimensionlessV1}), the position and dimensional
dependence of the potential energy are such that
\begin{equation}
V(\mu {\bf r}, \mu=1)= \mu^{-2}
V({\bf r}, \mu)
\;  ,
\label{eq:potential}
\end{equation}
for arbitrary $\mu$. In general, the function $V({\bf r}, \mu)$
may have a nontrivial (i.e., not quadratic) dependence with
respect to the parameter $\mu$, as displayed in
Eq.~(\ref{eq:Euler1}). However, those potentials devoid of
explicit dimensional scales are independent of any dimensional
parameter $\mu$; then their functional dependence is of the form
$V=V({\bf r})$ rather than of the form $V=V({\bf r},\mu)$. We can
also say that the second argument in $V=V({\bf r},\mu)$ is
actually spurious. An explicit mathematical statement of this
independence is
\begin{equation}
\frac{\partial}{\partial \mu} V({\bf r}, \mu) = 0 \;  .
\label{eq:scale_invariant}
\end{equation}
Even though this may be taken as the
primary definition,
it is convenient to derive a more illuminating form
by just eliminating the spurious $\mu$ dependence in
Eq.~(\ref{eq:potential}), i.e.,
\begin{equation}
V(\mu {\bf r})= \mu^{-2}
V({\bf r})
\;  .
\label{eq:homogpotential}
\end{equation}
Equation~(\ref{eq:homogpotential}), valid for $\mu>0$
(Eq.~(\ref{eq:positivity})), is identical to our earlier
homogeneous property~(\ref{eq:homogpotential_action}). An
alternative derivation of this remarkably simple property can be
obtained directly from Eq.~(\ref{eq:Euler1}) or by differentiation
of Eq.~(\ref{eq:potential}) with respect to $\mu$; then (after
setting the arbitrary scale equal to unity),
\begin{equation}
{\bf r} \cdot \mbox{\boldmath $\nabla$} V({\bf r}) = - 2 V({\bf r})
\; ,
\label{eq:Euler2}
\end{equation}
a relation that amounts to Euler's theorem for a  homogeneous
function of degree $-2$. In conclusion, the classes of
scale-invariant potentials and potentials without any explicit
dimensional scale are identical.

So far our discussion has only focused on the dimensions but not
on the magnitude of the potentials. It is now due time to
introduce a dynamical coupling parameter $\lambda$ to characterize
the ``strength'' of a given potential, according to
\begin{equation}
V({\bf r})= - \lambda  \,  W({\bf r})
\;  ,
\label{eq:W}
\end{equation}
where $W({\bf r})$ is a homogeneous function of degree $-2$.

The homogeneity displayed by Eq.~(\ref{eq:homogpotential})
has a straightforward consequence on the position dependence of
scale-invariant potentials. In effect, writing ${\bf r} = r
\hat{\bf r}$, where $\hat{\bf r}$ is the unit position vector, we
conclude that
\begin{equation}
V({\bf r})
= r^{-2} \,
V(  \widehat{\bf r} )
\;  ,
\label{eq:TP_general_form}
\end{equation}
so that the general scale-invariant potential has either one of
the following two forms:

(i) A generalized inverse square potential in any number $D$ of
dimensions,
\begin{equation}
V({\bf r})
= - \lambda
\,
\frac{
v
 ( \Omega^{({D})}  )}{ r^{2} }
\;  ,
\end{equation}
where the dimensionless
function $v (\Omega^{({D})} )$ explicitly depends upon
the $D$-dimensional solid angle $ \Omega^{({D})}
 \equiv  {\bf \widehat{ r}} $. In particular,
 $v=1 $ corresponds to the ordinary inverse
square potential~\cite{mot:49,cas:50,lan:77,mor:53,jac:72,gup:93},
\begin{equation}
V({\bf r})= -
\,
\frac{ \lambda }{ r^{2} }
\; ,
\label{eq:ISP_def}
\end{equation}
and $v =  \cos \theta$ amounts to  the dipole potential~\cite{lev:67},
\begin{equation}
V({\bf r})= - \lambda \,
\frac{\cos \theta}{ r^{2}}
\; ,
\label{eq:dipole_def}
\end{equation}
where $\theta$ is the polar angle (measured from the
orientation of the dipole moment)
and $\lambda $ is proportional to the magnitude of the dipole
moment.

(ii) A homogeneous pseudopotential of degree $-2$, for which
the most general form is
\begin{equation}
V({\bf r}) =
 - \lambda  \,
 r^{D-2} \, \delta^{(D)} ( {\bf r} )
 \; .
 \label{eq:pseudopotentials}
\end{equation}
It should be noticed that Eq.~(\ref{eq:pseudopotentials}) can be
transformed into a number of alternative forms involving the
radial delta function; in particular, it is equivalent to
 $\delta (r)/r $ and $\delta^{\, \prime} (r)$~\cite{pseudopotentials}.
 In this class, the two-dimensional delta-function potential
\begin{equation}
V({\bf r})= - \lambda  \,  \delta^{(2)} ({\bf r})
 \;
\label{eq:2D_delta_def}
\end{equation}
is the best known example, which we will analyze in
Section~\ref{sec:delta} in order to illustrate our general theory.

Equation~(\ref{eq:TP_general_form}) provides the limiting form of
the scale-invariant potential at infinity,
\begin{equation}
 \lim_{r \rightarrow \infty}
V({\bf r}) = 0
\;
\label{eq:TP_asymptotic}
\end{equation}
(which is identically true for pseudopotentials, for any $r \neq
0$). Equation~(\ref{eq:TP_asymptotic}) then implies that all
states with $E\geq 0$ are scattering states for any
scale-invariant potential. The sign of $\lambda$ in
Eqs.~(\ref{eq:W})--(\ref{eq:2D_delta_def}) has been chosen so that
$\lambda >0$ corresponds to an {\em attractive\/} potential,
wherever this concept is applicable.

\subsection{Scale Invariance and Eigenvalue Spectrum}
\label{sec:scaleinv}

The Hamiltonian $H_{\bf r}$ associated with a scale-invariant
potential,
\begin{equation}
{H}_{\bf r}
 = - \nabla^2 - \lambda \,
 W ({\bf r})
\;  ,
\end{equation}
is homogeneous of degree $-2$ with respect to the position
vector, that is,
\begin{equation}
{H}_{\zeta \, \bf r}= \zeta^{-2} {H}_{\bf r}
\;
\label{eq:scaleinv}
\end{equation}
for arbitrary $\zeta$, because both the kinetic and potential
energies have the same property. From our analysis of the last
section, this is another way of saying that
\begin{equation}
{\mathcal H}_{ { \mbox{\boldmath $ \scriptstyle{\xi}$}}   }
= \mu^{-2} H_{\mbox{\boldmath $\scriptstyle{ \xi}$}
/\mu }=
- \nabla_{ \mbox{ \boldmath $\scriptstyle{ \xi}$} }^2
- \lambda \, {\mathcal W}
( \mbox{\boldmath $ \xi$})
\;  ,
\label{eq:dimensionless_H}
\end{equation}
where
\begin{equation}
 {\mathcal W}
( \mbox{\boldmath $ \xi$})
= \mu^{-2} \, W ( \mbox{\boldmath $ \xi$}/\mu )
\;
\end{equation}
(cf.\ Eq.~(\ref{eq:dimensionlessV2})), is a dimensionless
Hamiltonian totally independent of any explicit dimensional scale.

A straightforward consequence of the scale invariance implied by
Eq.~(\ref{eq:scaleinv}) is the breaking of the discrete character
of the bound-state energy spectrum for attractive potentials. Let
us now show this. First, the
Hamiltonian~(\ref{eq:dimensionless_H}) associated with a
scale-invariant potential is an example of a local operator. As is
well known, the representative ${A}_{\bf r}$ of an abstract  local
operator ${A}$ is defined via
\begin{equation}
{A}_{\bf r} \delta^{({D})}
  ({\bf r}-{\bf r'}) =
\left\langle {\bf r} |A| {\bf r'} \right\rangle
\;  ,
\end{equation}
which implies that
\begin{equation}
{A}_{\bf r} \psi ({\bf r}) =
\left\langle {\bf r} |A| \psi \right\rangle
\;  .
\end{equation}
Equation~(\ref{eq:scaleinv}) refers to a particular case of a
local operator with scale dimension $s$, defined via
\begin{equation}
{A}_{\zeta \, \bf r} = \zeta^{-s} \,
{A}_{\bf r}
\;  ,
\end{equation}
for some exponent $s$ and for arbitrary scalar $\zeta>0$. With
arguments similar to those of Subsection~\ref{sec:scale_invariant2}, it
is easy to see that $s= {\rm dim} \left[ A_{\bf r} \right]$. Thus,
the Hamiltonian of scale-invariant potentials is a local operator
with scale dimension $s=2$.

The statement we wish to prove is that, for any local operator
with scale dimension $s$, the spectrum can only be continuous. In
effect, if $\psi_{a} ({\bf r})$ is an eigenfunction with
eigenvalue $a$,
\begin{equation}
{A}_{\bf r}
\psi_{a} ({\bf r})
= a  \,   \psi_{a} ({\bf r})
\;  ,
\end{equation}
then the rescaling
\begin{equation}
{\bf r}'= \zeta \, {\bf r}
\;
\end{equation}
implies that
\begin{equation}
{A}_{\bf r}
\psi_{a} (\zeta {\bf r})
= \zeta^{s} a  \,   \psi_{a} (\zeta {\bf r})
\;  ,
\end{equation}
whence
$\psi_{a} (\zeta {\bf r})$
is an eigenfunction of the same operator with eigenvalue $\zeta^{s} a$.
As this should be so for arbitrary $\zeta$, one concludes that:

(i)
If $a$ is a finite eigenvalue, then all
 real numbers of the same sign are eigenvalues.

(ii)
The eigenvalue spectrum is continuous.

Of course, this is what should be expected on intuitive grounds,
because if the spectrum were discrete then one would be
able to identify preferential
scales where none is defined a priori.
For example,
the rescaling ${\bf r}'= \zeta {\bf r}$
for the plane-wave eigenstates  $e^{i {\bf k} \cdot {\bf r}}$
of the scale-invariant free-particle Hamiltonian
amounts to a rescaling of the corresponding momentum
${\bf k}'= \zeta {\bf k}$, whence all positive energies
$E=k'^{2}=\zeta^2 k^2$
can be reached
by continuously varying the parameter $\zeta$.

Therefore, for the Hamiltonian of attractive scale-invariant
potentials, the corresponding implications  are

(i)
The energy spectrum is
either not bounded from below
or, if it is, it can only start at $E=0$ and be of the scattering 
type.

(ii)
The bound-state energy spectrum, if it exists,
is continuous.

Thus, it is clear that, for a given unregularized scale-invariant
potential, in addition to the continuous scattering spectrum with
energies $0 \leq E < \infty$ (as required by
Eq.~(\ref{eq:TP_asymptotic})), there are only three possibilities
for the bound-state spectrum:
\begin{enumerate}
\item
Spectrum devoid of bound states.
\item
Continuous bound-state spectrum with energies between
$E=-\infty$ and $E=0$.
\item
Singular bound-state spectrum with a unique energy level at
$E=-\infty$.
\end{enumerate}
The fact that no other cases are possible is just a consequence of
the scale symmetry.

In other words, in the first category the potential is so ``weak''
that if fails to generate any bound states; in fact, this
situation is familiar: it is characteristic of any repulsive
potential. The subtlety lies in that a ``weak'' attractive
potential behaves in all respects as a repulsive one: it only has
a continuous scattering spectrum extending from $E=0$ to
$E=\infty$. As an alternative, the potential may be so ``strong''
that it produces an example of the second category, where it
breaks down both the existence of a lower bound and the discrete
nature of the bound-state spectrum. It is well known that both
cases (categories 1 and 2) are realized by the inverse square
potential with weak and strong coupling,
respectively~\cite{mot:49,cas:50,lan:77,mor:53,jac:72,gup:93}.
Yet, the singular case of category 3 provides an alternative for a
``strong'' potential---a behavior exhibited by the two-dimensional
delta-function potential~\cite{hua:82,tho:79,jac:91}.

With regard to scattering, one may use the fact that
the theory is scale-invariant and that the poles of the scattering 
matrix
on the imaginary energy axis correspond to the bound states.
These facts imply that:

(i)
When there are no bound states, the scattering matrix has no poles
and is manifestly scale-invariant, i.e., independent of the energy.

(ii)
When the spectrum is singular (either categories 2 or 3),
the scattering matrix exhibits the corresponding singular behavior.

A remark about the need for renormalization is in order. A theory
that produces no bound states and a scale-invariant S-matrix
(category 1) needs no regularization. In effect, such theory
displays no divergence whatsoever and, as we have seen, its
spectrum is identical in every respect to that of repulsive
potentials, so that scale-invariance is maintained even in the
quantum-mechanical theory. Instead, regularization and
renormalization are needed for cases 2 and 3 above, an issue to
which we now turn our attention.

\section{DIMENSIONAL TRANSMUTATION VIA DIMENSIONAL REGULARIZATION}
\label{sec:DT_via_DR}

\subsection{Dimensional Regularization of Scale-Invariant Potentials}
\label{sec:drtp}

Let us consider a scale-invariant potential $V({\bf r})$ in
$D_{0}$ dimensions. The corresponding $D_{0}$-dimensional
Schr\"{o}dinger equation is
\begin{equation}
\left[ - \nabla_{{\bf r}, {{D_{0}}} }^2
+ V({\bf r}) \right]
\,  \Psi ({\bf r})
=
E \,  \Psi ({\bf r})
\;  ,
\label{eq:Schr1nu0}
\end{equation}
where $\nabla_{{\bf r},
{{D_{0}}}}^2 $
is the $D_{0}$-dimensional Laplacian.

We have seen in Subsection~\ref{sec:scaleinv} that if $V({\bf r}) $ is
of the scale-invariant type, then its unregularized bound-state
spectrum is either nonexistent or not bounded from below.
Therefore, the difficulty here resides in  that,
 in the initial dimension
$D_{0}$, the problem is singular and has to be
regularized.

In this paper we will use dimensional regularization,
a technique originally developed for quantum field
theory~\cite{bol:72,lei:75}
and which we now adapt to nonrelativistic quantum mechanics.
The $D$-dimensional generalization of Eq.~(\ref{eq:Schr1nu0})
is of the form
\begin{equation}
\left[ - \nabla_{{\bf r},
{{D}} }^2
+ V^{({D})}({\bf r}) \right]
\,  \Psi ({\bf r})
=
E \,  \Psi ({\bf r})
\;  ,
\label{eq:Schr1nu1}
\end{equation}
where
\begin{equation}
V^{({D})}({\bf r}) =
- \lambda_{B}
\,  W^{({D})}({\bf r})
\;  ,
\end{equation}
with $ \lambda_{B}$ being the bare coupling constant (see
Subsection~\ref{sec:coupling}), while $ W^{({D})}({\bf r}) $
is an appropriate generalization to $D$ dimensions of
the original $D_{0}$-dimensional potential,
with the only constraint
\begin{equation}
\lim_{{D} \rightarrow {D_{0}}}
W^{({D})}({\bf r}) =
W({\bf r})
\;  ,
\end{equation}
and with  $W({\bf r}) $ defined by Eq.~(\ref{eq:W}).

Of course, the whole purpose of this regularization is to produce
a scenario where the generalized potential is no longer
scale-invariant in a dimension close but not equal to $D_{0}$. In
other words, we require that, for arbitrary $\epsilon \neq 0$,
  $V^{(
{{D_{0}}} - \epsilon)}({\bf r})$
not be a homogeneous function of degree $-2$.

Even though the requirement above allows infinitely many possible
generalizations, a particularly simple prescription can be
developed by the use of Fourier transforms, just like it is done
in  quantum field theory. It turns out that a simple property of
Fourier analysis immediately suggests the generalization: if
$f({\bf r})$ is a $D$-dimensional homogeneous function of degree
$\beta$, then its Fourier transform $\widetilde{f}({\bf k})$ is
homogeneous of degree $-(D + \beta)$. With this property in mind,
we define the Fourier transform of the original scale-invariant
potential,
\begin{equation}
\widetilde{W}({\bf k})
= {\mathcal F}_{( { {D_{0}}}
)} \left\{
W({\bf r}) \right\}
=
\int d^{{D_{0}}} r \; e^{-i{\bf k} \cdot {\bf r}}
\, W({\bf r})
\;  ,
\label{eq:fourier}
\end{equation}
which we analytically continue to $D$ dimensions
with the prescription that, in Fourier space, the $D$-dimensional
functional form should
be the same as the $D_{0}$-dimensional functional form, i.e.,
\begin{equation}
    \widetilde{W}^{({D})}({\bf k}) =
              \widetilde{W}({\bf k})
\;  .
\end{equation}
Finally a $D$-dimensional inverse Fourier transform
$  {\mathcal F}^{-1}_{({D}) }$
provides the desired generalization
in the position representation,
i.e,
\begin{equation}
W^{({D})}({\bf r})
= {\mathcal F}_{({D})}^{-1} \left\{
\widetilde{W}({\bf k})  \right\}
=
\int \frac{ d^{{D}} k }{(2\pi)^{{D}}}
\; e^{i{\bf k} \cdot {\bf r}}
\, \widetilde{W}({\bf k})
\;  ,
\label{eq:inversefourier}
\end{equation}
where all the vectors are now $D$-dimensional.

The process represented by Eqs.~(\ref{eq:fourier})
and (\ref{eq:inversefourier})
involves a  dimensional continuation that
can be summarized in the following succinct expression
for the potential
\begin{equation}
W^{({D})}({\bf r}_{{D}}) =
\int \frac{ d^{{D}}
k_{ { {D}} }}
{(2\pi)^{{D}}}
   \; e^{i{\bf k}_{{D}}
\cdot {\bf r}_{{D}} }
\,
\left[
\int d^{{D_{0}}}
r_{ {{D_{0}}}}
    \; e^{-i{\bf k}_{{D_{0}}}
\cdot {\bf r}_{{D_{0}}}}
\, W({\bf r}_{{D_{0}}} )
\right]_{ {\bf k}_{{D_{0}}}
\rightarrow {\bf k}_{{D}} }
\;  ,
\label{eq:dimcontinuation}
\end{equation}
where the subscripts $D$ and $D_{0}$ explicitly indicate the
dimension of the corresponding vector and the symbol
${ {\bf k}_{{D_{0}}}
\rightarrow {\bf k}_{{D}} }$
stands for the dimensional ``jump'' in momentum space that defines
the dimensionally continued potential.
The whole process can be represented
by means of  the commutative diagram
\begin{equation}
\diagram{
& \mbox{\rm real space} & & \mbox{\rm reciprocal space} \cr
\mbox{$D_{0}$  {\rm dimensions}:}
        &  W({\bf r}) = W^{(
           { {D_{0}}}  )}({\bf r})
  &  \diagramrightarrow{
       \mbox{${\mathcal F}_{(  { {D_{0}}} )}$}  }{}
          &  \widetilde{W}( {\bf k}) =
           \widetilde{W}^{({{D_{0}}} )}({\bf k})
    \cr
 &  \diagramdownarrow{
        \mbox{${\mathcal D}_{
          {{D_{0} }} \rightarrow
    {{D}}  }$} }{} & &
 \diagramdownarrow{ \mbox{${\mathcal D}_{
                  {{D_{0}}}
          \rightarrow  {{D}} }$} }{}  \cr
\mbox{$D$  {\rm dimensions}: }  &
               W^{({D})}({\bf r})   &
     \diagramleftarrow{
       \mbox{${\mathcal F}^{-1}_{({D})}$}  }{}
         &
\widetilde{W}^{({D})}({\bf k}) =
              \widetilde{W}({\bf k})
    \cr 
}
\;  ,
\label{eq:diagramdef1}
\end{equation}
where ${{\mathcal D}_{
     {{D_{0}}} \rightarrow
  {{D}}  }}$ is a shorthand
for dimensional continuation from $D_{0}$ to $D$ dimensions.
Correspondingly, the degree of homogeneity of a scale-invariant
potential is transformed according to
\begin{equation}
\diagram{ \stackrel{ \mbox{$W({\bf r})$}  }{ {\rm degree}= -2}
  &  \diagramrightarrow{ \mbox{${\mathcal F}_{(
              {{D_{0}}} )}$}  }{}
          &
  \stackrel{ \mbox{$\widetilde{W}({\bf k})$}  }{
                   {\rm degree}= 2 - D_{0} } \cr
 \diagramdownarrow{ \mbox{${\mathcal D}_{ {{D_{0}}}
                   \rightarrow  {{D}} }$} }{}  & &
 \diagramdownarrow{ \mbox{${\mathcal D}_{
             {{D_{0}}}
            \rightarrow  {{D}} }$} }{}  \cr
\stackrel{
\mbox{$W^{({D})}({\bf r})$} }{
{\rm degree}= -2+ \epsilon } &
  \diagramleftarrow{ \mbox{${\mathcal F}^{-1}_{({D})}$}  }{}
    &   \stackrel{
          \mbox{$\widetilde{W}^{({D})}({\bf k})$}  }{
                 {\rm degree}= 2 - D_{0}} \cr
}
\;  \;  ,
\label{eq:diagramdef2}
\end{equation}
where
\begin{equation}
\epsilon = D_{0}- D
\;  .
\end{equation}
Diagram~(\ref{eq:diagramdef2}) explicitly shows that the
$D$-dimensional real-space continuation of the potential is {\em
not\/} of the scale-invariant type, because its degree of
homogeneity is $-2+\epsilon$ (with $\epsilon \neq 0$),  rather
than $-2$.

For example, when the criteria above
are applied to the two-dimensional delta-function and
inverse square potentials,
one obtains the dimensional continuations summarized in
Table~\ref{tab:dimcontinuation}~\cite{cam:00b_remark}.

\begin{center}
\begin{tab}
\hspace{-.05in}. \hspace{.1in}
{\rm Dimensional Continuations $W^{({D})}({\bf r})$
for the Two-Dimensional Delta-Function
and Inverse Square Potentials.}
\label{tab:dimcontinuation}
\end{tab}
\end{center}

\begin{center}
\begin{tabular}{ccc}  \hline\hline
{ Potential $W({\bf r})$} &  \hspace{.05in} Dimension $D_{0}$
& \hspace{.05in}
 {Dimensional continuation $W^{({D})}({\bf r})$}
\\ \hline
$ \delta^{(2)}({\bf r})$ & 2 &
 $\left. \delta^{({D})}({\bf r}) \right|_{D=2-\epsilon} $  \\
 $ r^{-2}$  & Arbitrary &
  $\pi^{\epsilon/2} \, \Gamma(1-\epsilon/2) \,
r^{-(2-\epsilon)}$
\\
\hline\hline
\end{tabular}
\end{center}

\subsection{Dimensional Transmutation of the Coupling Parameter}
\label{sec:coupling}

In the analysis above
we have defined the appropriate functional dependence of the
dimensionally continued potential
but we have not spelled out the dimensionality change experienced
by the coupling parameter.
This change occurs in Eq.~(\ref{eq:dimcontinuation}),
when the dimensional ``jump''
${ {\bf k}_{{D_{0}}}
\rightarrow {\bf k}_{{D}} }$
is performed, according to
\begin{equation}
\left[ W^{({D})}
({\bf r}_{{D}}) \right]
= \Lambda^{D-D_{0}}
 \left[ W^{( {D_{0}} )}
({\bf r}_{{D_{0}}}) \right]
\; .
\label{eq:dimensional_creation}
\end{equation}
Equation~(\ref{eq:dimensional_creation})
amounts to the ``creation'' of a physical dimension
$L^{-({D} - {D}_{0})}
=L^{\epsilon}$.
If the physical dimensions of the potential energy are to remain
unchanged, then the bare coupling constant
$\lambda_{B}$ should acquire
the physical dimensions $L^{-\epsilon}$, that is,
its dimensionality should become
${\rm dim} \left[ \lambda_{B} \right] = \epsilon$.
A convenient way of parametrizing this dimensionality change is by
the introduction of an {\em arbitrary\/}
 inverse-length scale $\mu$, i.e.,
by the replacement
\begin{equation}
\lambda \rightarrow
\lambda_{B}
=
\lambda \mu^{\epsilon}
\;  ,
\label{eq:dimensionlesscoupling}
\end{equation}
where $\lambda$ is dimensionless.
When this dimensionality
change is made explicit in Eq.~(\ref{eq:dimcontinuation}),
we obtain a dimensionally continued potential
\begin{eqnarray}
V^{({D})}({\bf r})  & = &
- \lambda \,
\mu^{\epsilon}
\,
W^{({D})}({\bf r})
\;  ,
\label{eq:dimcontinuation2}
\\
& = &  -
\lambda
 \mu^{\epsilon}  \,
\int \frac{ d^{ {D}}
k_{ {D} }}
{(2\pi)^{{D}}}
   \; e^{i{\bf k}_{{D}}
\cdot {\bf r}_{{D}} }
\,
\left[
\int d^{ {D_{0}}}
r_{{D_{0}}}
    \; e^{-i{\bf k}_{{D_{0}}}
\cdot {\bf r}_{{D_{0}}}}
\, V({\bf r}_{{D_{0}}} )
\right]_{ {\bf k}_{{D_{0}}}
\rightarrow {\bf k}_{{D}} }
\;  ,
\label{eq:dimcontinuation3}
\end{eqnarray}
i.e., the dimensional jump is performed simultaneously
with the introduction of the factor $\mu^{\epsilon}$.
Notice that
Eqs.~(\ref{eq:dimcontinuation}) and~(\ref{eq:dimcontinuation3}),
as well as diagram (\ref{eq:diagramdef2})
show that even though
\begin{equation}
{\rm dim} \left[  V^{({D})} ({\bf r}) \right] =2
\;  ,
\end{equation}
in fact
\begin{equation}
{\rm dim} \left[  W^{({D})} ({\bf r}) \right] =2 - \epsilon
\;  .
\end{equation}
Thus, one can write
\begin{equation}
W^{({D})} ({\bf r})  =
\mu^{2 - \epsilon}
\,  {\mathcal W}^{({D})} (\mu {\bf r})
\;  ,
\label{eq:dimensionless_W}
\end{equation}
where $ {\mathcal W}^{({D})}( \mbox{\boldmath $ \xi$} )  $
is the dimensionless counterpart of
$W^{({D})} ({\bf r})$; then, from
Eqs.~(\ref{eq:dimensionlessV2}),
(\ref{eq:dimcontinuation2}), and (\ref{eq:dimensionless_W}),
\begin{equation}
 {\mathcal V}^{({D})}( \mbox{\boldmath $ \xi$} )  =
- \lambda
\,
 {\mathcal W}^{({D})}( \mbox{\boldmath $ \xi$} )
\;  .
\end{equation}

The dimensionality change represented by 
Eq.~(\ref{eq:dimensionlesscoupling})
introduces a completely arbitrary dimensional scale $\mu$.
The  replacement
of a dimensionless coupling constant by an arbitrary dimensional scale
is the phenomenon of dimensional transmutation seen from the
dimensional-regularization viewpoint.
In addition,
Eq.~(\ref{eq:dimensionlesscoupling}) indicates that the space dimension
$D$ plays a pivotal role
 in the determination of the physical dimensions
of the coupling constant.

Equations~(\ref{eq:Schr1nu1})
and (\ref{eq:dimcontinuation2})
imply that the dimensionally regularized Schr\"odinger
equation has the explicit form
\begin{equation}
\left[ - \nabla_{{\bf r}, {D} }^2
- \lambda \mu^{\epsilon}  \,
W^{({D})}({\bf r}) \right]
\,  \Psi ({\bf r})
=
E \,  \Psi ({\bf r})
\;  ,
\label{eq:Schr1nu2}
\end{equation}
where $W^{({D})}({\bf r}) $
is a homogeneous function of degree
$-2+\epsilon$.
Ultimately, in order to make the potential less singular, it
is necessary to choose $\epsilon>0$ (that is, $D<D_{0}$),
whence proper regularization is achieved in the limit
\begin{equation}
\epsilon=0^{+}
\;  .
\label{eq:epsilon_limit}
\end{equation}

Alternatively, the dimensionally regularized
Schr\"odinger equation~(\ref{eq:Schr1nu2})
can be rewritten in the dimensionless form
of Eq.~(\ref{eq:Schr2}), i.e.,
\begin{equation}
\left[ -
\nabla_{
\mbox{\boldmath $\scriptstyle{ \xi}$},
{D} }^2   -
    \lambda \,  {\mathcal W}^{({D})}
( \mbox{\boldmath $ \xi$} )  \right]
\,  \Phi (  \mbox{\boldmath $ \xi$} )
= \eta
 \, \Phi (  \mbox{\boldmath $ \xi$} )
\;  .
\label{eq:Schr2nu1}
\end{equation}
If the problem  posed by Eq.~(\ref{eq:Schr2nu1})
with the limit~(\ref{eq:epsilon_limit})
is {\em regular\/}, then
its solution provides regular eigenfunctions
 $\Phi (  \mbox{\boldmath $ \xi$} )$
corresponding to the eigenvalues $\eta$. These eigenvalues depend
upon the dimensionless parameters $\epsilon$ and $\lambda$. Unlike
the dependence of $\eta$ on  $\epsilon$, which is
potential-dependent, the dependence of $\eta$ on $\lambda$ is the
{\em same\/} for all scale-invariant potentials. This can be
understood as follows:
\begin{enumerate}
\item
The Schr\"odinger equation~(\ref{eq:Schr2nu1})
is dimensionless and independent of $\mu$.
\item
Instead, the
Schr\"odinger equation~(\ref{eq:Schr1nu2}) is
more explicit in that it displays the dimensional scales
$E$ and $\mu$.
In addition, $\lambda$ and $\mu$ appear in it
{\em only\/} in its second term
and through the bare coupling
constant $\lambda_{B}$, Eq.~(\ref{eq:dimensionlesscoupling}).
This can be made more explicit by rewriting
Eq.~(\ref{eq:Schr1nu2}) in terms of
$\lambda_{B}$, i.e.,
\begin{equation}
\left[ - \nabla_{{\bf r}, {D} }^2
- \lambda_{B}   \,
W^{({D})}({\bf r}) \right]
\,  \Psi ({\bf r})
=
E \,  \Psi ({\bf r})
\;  .
\label{eq:Schr1nu3}
\end{equation}
\item
Equation~(\ref{eq:dimensionlesscoupling})
is the basis for the introduction of
an effective inverse length
\begin{equation}
\widehat{\mu}
=
\lambda_{B}^{1/\epsilon}
=
\lambda^{1/\epsilon} \, \mu
\; ,
\label{eq:widehat_mu}
\end{equation}
where $\lambda>0$ is assumed because we only
need to regularize the potential when it is attractive.
Equation~(\ref{eq:widehat_mu})
allows for the rescaling of the energy
\begin{equation}
\widehat{\eta}
=
\widehat{\mu}^{-2}  \,
E
=
 \lambda^{-2/\epsilon} \,
 \eta
\;  ,
\label{eq:widehat_eta}
\end{equation}
of the potential energy
\begin{equation}
\widehat{  {\mathcal W} }
=
\widehat{\mu}^{-(2-\epsilon)} \, W
=
\lambda^{-2/\epsilon + 1} \,  {\mathcal W}
\;  ,
\label{eq:widehat_W}
\end{equation}
as well as of the dimensionless position
\begin{equation}
\widehat{\mbox{\boldmath $ \xi$}}
 =
 \widehat{\mu} \, {\bf r}
= \lambda^{1/\epsilon} \, {\mbox{\boldmath $ \xi$}}
\;  .
\end{equation}
Then, the Schr\"odinger
equation~(\ref{eq:Schr1nu3}) takes
the form
\begin{equation}
\left[ - \nabla_{
\widehat{\mbox{\boldmath $\scriptstyle{\xi}$}}, D }^2   -
   \widehat{ {\mathcal W} }^{({D})}
( \widehat{ \mbox{\boldmath $ \xi$}} )
\right]
\,  \widehat{\Phi} (  \widehat{\mbox{\boldmath $ \xi$}} )
= \widehat{\eta}
 \, \widehat{\Phi} ( \widehat{ \mbox{\boldmath $ \xi$}} )
\;  .
\label{eq:Schr2nu2}
\end{equation}
\end{enumerate}

Equations~(\ref{eq:Schr1nu2}), (\ref{eq:Schr2nu1}), and
(\ref{eq:Schr2nu2}) are equivalent before taking the limit
$\epsilon=0^{+}$. When actually solving the Schr\"odinger equation
for a specific potential, it is more straightforward to use
(\ref{eq:Schr1nu2}) because it makes all the variables explicit,
while (\ref{eq:Schr2nu1}) is just a convenient way of relating the
dimensional scales from scratch. On the other hand, at the
conceptual level, Eq.~(\ref{eq:Schr2nu2}) provides the
``universal'' connection between $\lambda$ and $\eta$ for all
scale-invariant potentials. In other words,
Eq.~(\ref{eq:widehat_eta}) defines the relationship between the
mathematical eigenvalues $\widehat{\eta}$ of
Eq.~(\ref{eq:Schr2nu2}) and the physical or dimensional
eigenvalues $E$ of Eq.~(\ref{eq:Schr1nu2}), which are explicit
functions of the parameters $\mu$ and $\lambda$. A convenient form
of this ``universal'' condition satisfied by the energy
eigenvalues of Eq.~(\ref{eq:Schr2nu2}) reads (from
Eq.~(\ref{eq:widehat_eta}))
\begin{equation}
 \lambda \,
\Xi (\epsilon)
\, | \eta (\epsilon) |^{-\epsilon/2} = 1
\; ,
\label{eq:eigenvalues_general}
\end{equation}
which, in terms of dimensional variables,
explicitly states that
\begin{equation}
 \lambda \, \mu^{\epsilon} \,
\Xi (\epsilon)
\, | E  (\epsilon) |^{-\epsilon/2} = 1
 \; .
\label{eq:eigenvalues_general2}
\end{equation}
Any of the Eqs.~(\ref{eq:eigenvalues_general})
and (\ref{eq:eigenvalues_general2})
would be referred to as the ``master eigenvalue equation,''
which provides the required energies if
the  mathematical function
\begin{equation}
 \Xi (\epsilon)
= \left|
\widehat{\eta} \right|^{\epsilon/2}
\geq  0
\;
\label{eq:master_eigenvalue_function}
\end{equation}
is known.  In fact, Eq.~(\ref{eq:master_eigenvalue_function})
shows that $\Xi (\epsilon) $
is completely determined by the functional form of the potential
$\widehat{ {\mathcal W} }^{({D})}
( \widehat{\mbox{\boldmath $ \xi$}} ) $,
through the solution of the differential equation~(\ref{eq:Schr2nu2}).
We shall refer to $\Xi (\epsilon) $
as the ``energy generating function.''

In this paper, we will exemplify the regularization
procedure and computation of the
energy generating function for
the two-dimensional delta-function potential. A
more detailed discussion of this potential, as well as the
corresponding treatment of the inverse square
potential, can be found in the second paper of this
series~\cite{cam:00b}.

\section{RENORMALIZATION OF SCALE-INVARIANT POTENTIALS}
\label{sec:renormalization}

\subsection{Regularized Bound-State Sector}
\label{sec:BS_sector_regularized}

The theory of Subsection~\ref{sec:coupling}
applies equally well to both bound and scattering states.
In both cases, however,
we may assume an attractive potential $\lambda>0$,
which is the type that possibly requires renormalization.

Let us now consider the bound-state sector,
for which Eq.~(\ref{eq:Schr2nu1})
provides  a discrete sequence of energy
eigenvalues $\eta_{n}$, in the regularized version of the theory.
These eigenvalues explicitly depend upon
the discrete set of quantum numbers
 $n=\left(n_{1}, \dots , n_{D} \right)$,
ordered as an increasing sequence
in such a way that, for sufficiently small $\epsilon$,
$E_{n}(\epsilon) \leq E_{ n^{\prime} } (\epsilon) $
if $n_{j} \leq n^{\prime}_{j}$ for all $j=1, \ldots , D$
(additional ordering rules are needed
in the presence of degeneracies, but they are immaterial to our 
discussion).
In particular, the ground state will be labeled with the lowest numbers
of the sequence.

Our goal is to extract additional information from
Eq.~(\ref{eq:eigenvalues_general}), which we now rewrite
\begin{equation}
 \lambda \,
\Xi_{n} (\epsilon) \, | \eta_{n} (\epsilon) |^{-\epsilon/2} = 1
\; ,
\label{eq:eigenvalues_BS}
\end{equation}
with $ \Xi_{n}(\epsilon) = \left|
\widehat{\eta}_{n} \right|^{\epsilon/2} $.
This can be accomplished by
defining the variables
\begin{equation}
\lambda^{(\ast)}_{n} = \left[
\lim_{\epsilon\rightarrow 0}
\Xi_{n}(\epsilon)
\right]^{-1}
= \lim_{\epsilon\rightarrow
0} \left| \widehat{\eta}_{n}(\epsilon)
\right|^{-\epsilon/2}
\;  ,
\label{eq:lambda_star}
\end{equation}
such that
\begin{equation}
\lim_{\epsilon\rightarrow 0}
 \left|\eta_{n}(\epsilon)\right| =
\lim_{\epsilon\rightarrow 0} \left[\frac{\lambda}
{\lambda^{(\ast)}_{n}} \right]^{2/\epsilon} \;  .
\label{eq:limit_dimensionless_energy}
\end{equation}
In Eq.~(\ref{eq:limit_dimensionless_energy})
one can see that the unregularized energy is critically dependent
on the ratio $\lambda/ \lambda^{(\ast)}_{n}$,
in the limit $\epsilon \rightarrow 0$.
Thus, $ \lambda^{(\ast)}_{n}$ acts as
a critical coupling strength for the given
energy level labeled by $n$.
From now on, we will identify the following three regimes:

(i) Strong coupling,
characterized by $\lambda > \lambda^{(\ast)}_{n}$,
for which  Eq.~(\ref{eq:limit_dimensionless_energy})
gives a bound state at $-\infty$.

(ii) Weak coupling,
characterized by $\lambda < \lambda^{(\ast)}_{n}$,
for which  Eq.~(\ref{eq:limit_dimensionless_energy})
gets rid of the bound state by pushing it all the way up to  $0$.

(iii) Critical coupling,
characterized by $\lambda = \lambda^{(\ast)}_{n}$,
for which additional analysis is needed.

In fact, Eq.~(\ref{eq:limit_dimensionless_energy})
implies the following behavior according to the
values of the critical
coupling:

(a)  $ \lambda^{(\ast)}_{n}=0$ amounts to
a strong coupling  for all  finite and positive $\lambda$,
a condition that is manifested by the
``collapse'' of  the given bound state,
 $\eta_{n}\rightarrow -\infty$.

(b) $ \lambda^{(\ast)}_{n}=\infty$ amounts to a weak coupling
 for all finite $\lambda>0$,
a condition that is manifested by the loss of the regularized
bound state, i.e.,
 $\eta_{n}\rightarrow 0$.

(c) $ 0< \lambda^{(\ast)}_{n}< \infty$
permits the existence of the three possible regimes.

A few results are implied by the above analysis.
First, because of the assumed ordering of quantum numbers,
 from Eq.~(\ref{eq:lambda_star}),
it follows that
$ \lambda^{(\ast)}_{n}
\leq
 \lambda^{(\ast)}_{ n^{ \prime } }$,
when $n_{j} \leq n^{\prime}_{j}$ for all $j=1, \ldots , D$.
In particular, for the ground state,
which we will subsequently label with ${\rm (gs)}$,
we define the ``principal'' critical coupling $ \lambda^{(\ast)}$,
which satisfies the condition
\begin{equation}
 \lambda^{(\ast)}
=
 \lambda^{(\ast)}_{\rm (gs)}
\leq
 \lambda^{(\ast)}_{n}
\;  .
\label{eq:critical_inequality}
\end{equation}
For example, if the coupling is weak for the ground state,
it is also weak for all other states, so that the unregularized
theory is completely deprived from bound states.

The analysis above  assumes that $\lambda$
is independent of $\epsilon$ and displays the singular behavior 
associated with dimensional transmutation as $\epsilon \rightarrow 0$.
Renormalization is called for in order to obtain meaningful results.

\subsection{Renormalized Bound-State Sector}
\label{sec:BS_sector_renormalized}

Equation~(\ref{eq:eigenvalues_BS})
provides a regularization of the
energy levels in terms of the parameter $\epsilon$.
In this section we introduce the general strategy for
renormalization and use it to reach a few general conclusions
about dimensional transmutation.

In order to obtain finite results, it is necessary to renormalize
the energy levels by the following procedure:

(i) Letting
 the coupling constant $\lambda$ be a function of
the regularization parameter $\epsilon$, i.e.,
$ \lambda = \lambda (\epsilon)$.

(ii)
Adjusting  $\lambda (\epsilon)$  by comparison with a specific
bound state, which is conveniently chosen to be the ground
state of the theory; notice that if bound states exist at all,
the ground state
is the only one that is guaranteed to exist.
We will refer to this procedure as ground-state renormalization.

Consequently,
in the following analysis, it will prove useful
to compare the values of the function
$\Xi_{n}(\epsilon)$ with its ground-state value
$\Xi_{_{\rm (gs)}} (\epsilon)   $,
by means of the replacement
\begin{equation}
\Xi_{n}(\epsilon)
=
\Xi_{_{\rm (gs)}}  (\epsilon)
\,
\left[ 1 + \frac{\epsilon}{2}  \,
{\mathcal R}_{n} (\epsilon)   \right]
\;  ,
\label{eq:master_eigenvalue_function_remainder}
\end{equation}
which defines a new function ${\mathcal R}_{n}(\epsilon)$,
with the obvious implication that
\begin{equation}
{\mathcal R}_{_{\rm (gs)}}  (\epsilon) = 0
\;  .
\label{eq:master_eigenvalue_function_remainder_GS}
\end{equation}
Then, the regularized dependence of the energy levels
with respect to  $\epsilon$ can be
derived from (\ref{eq:eigenvalues_BS}) in the
limit~(\ref{eq:epsilon_limit}), and using
Eq.~(\ref{eq:master_eigenvalue_function_remainder}), which implies
\begin{eqnarray}
 | \eta_{n}
(\epsilon)
| &  =  &
\left[
\lambda (\epsilon)
\,
\Xi_{_{\rm (gs)}}  (\epsilon)
\right]^{2/\epsilon}
\,
\left[ 1 +  \frac{\epsilon}{2} \,  {\mathcal R}_{n} (\epsilon)
\right]^{2/\epsilon}
\label{eq:BS_energies_asymptotic1}
\\  & \approx &
\left[
\lambda (\epsilon)
\,
\Xi_{_{\rm (gs)}}
 (\epsilon)
\right]^{2/\epsilon}
\,
\exp \left[  {\mathcal R}_{n} (\epsilon)  \right]
\;   .
\label{eq:BS_energies_asymptotic2}
\end{eqnarray}

Furthermore,
the analysis of the previous section shows
that  finite results follow {\em only\/} if the
coupling constant takes a  critical value.
For the ground state, this requires that
 \begin{equation}
\lambda(\epsilon)
 \stackrel{\scriptscriptstyle (\epsilon \rightarrow 0)}{\sim}
\left[
\Xi_{_{\rm (gs)}} (\epsilon)
\right]^{-1}
\;  .
\label{eq:lambda_epsilon_asymptotic}
\end{equation}
Even though Eq.~(\ref{eq:lambda_epsilon_asymptotic}) is sufficient
for renormalization purposes, let
us consider a more general expression
\begin{equation}
\lambda (\epsilon)=
\left[
\,
\Xi_{_{\rm (gs)}} (\epsilon)
\right]^{-1}
\,
\left[ 1 +
\frac{\epsilon}{2}  \,
g (\epsilon)   \right]
\;  .
\label{eq:lambda_epsilon}
\end{equation}
Equation~(\ref{eq:lambda_epsilon})
defines a residual coupling function $g(\epsilon)$,
which---according to the definition of critical coupling,
Eq.~(\ref{eq:lambda_star})---should have the limiting behavior
\begin{equation}
\epsilon \,
g(\epsilon) = o(1)
\;  .
\end{equation}

As a consequence,
from Eqs.~(\ref{eq:BS_energies_asymptotic2}) and 
(\ref{eq:lambda_epsilon}),
\begin{equation}
 | \eta_{n}(\epsilon)  |  =
\exp \left[ g(\epsilon) + {\mathcal R}_{n} ( \epsilon) \right]
\;   ,
\label{eq:BS_energies_asymptotic3}
\end{equation}
and
\begin{equation}
 \left|
 \frac{\eta_{n}(\epsilon) }{ \eta_{_{\rm (gs)}}
(\epsilon)  }
  \right|
  =
\exp \left[  {\mathcal R}_{n} (\epsilon)
\right]
\;   .
\label{eq:excited_energies_asymptotic1}
\end{equation}
From the form of Eqs.~(\ref{eq:BS_energies_asymptotic3})
and (\ref{eq:excited_energies_asymptotic1}),
it proves convenient to resolve both ${\mathcal R}_{n}(\epsilon) $ 
and
$g(\epsilon)$ into their various
components, i.e.,
 \begin{equation}
{\mathcal R}_{n} (\epsilon) =
 {\mathcal R}_{n}^{(-)}(\epsilon)
+{\mathcal R}_{n}^{(0)}+{\mathcal R}_{n}^{(+)}(\epsilon)
\;
\end{equation}
and
 \begin{displaymath}
g(\epsilon) = g^{(-)}(\epsilon)
 +g^{(0)}+g^{(+)}(\epsilon)
\;  ,
\end{displaymath}
 where ${\mathcal R}_{n}^{(-)}(\epsilon)$ and $g^{(-)}(\epsilon)$ are
the divergent pieces;
 ${\mathcal R}_{n}^{(0)}$ and $g^{(0)}$ are
the limits, for $\epsilon=0^{+}$, of the finite parts;
and ${\mathcal R}_{n}^{(+)}(\epsilon), g^{(+)}(\epsilon) = o(1)$.
 Then Eq.~(\ref{eq:BS_energies_asymptotic3})
will again give 0 or $\infty$, unless
 \begin{equation}
 [g^{(-)}(\epsilon)+{\mathcal R}_{n}^{(-)}(\epsilon)]
+[g^{(0)}+{\mathcal R}_{n}^{(0)}]+
[g^{(+)}(\epsilon)
 +{\mathcal R}_{n}^{(+)}(\epsilon)]=O(1)
\;  .
\label{eq:g_constraint}
\end{equation}
From now on, the terms
$g^{(+)}(\epsilon)$ and $ {\mathcal R}_{n}^{(+)}(\epsilon)$
can and will be omitted, as they are clearly irrelevant
at the level of the renormalized energies.
In turn, in Eq.~(\ref{eq:g_constraint}),
 the terms $[g^{(-)}(\epsilon)+{\mathcal R}_{n}^{(-)}(\epsilon)]$
would give a divergent contribution unless
 \begin{equation}
g^{(-)}(\epsilon)=-{\mathcal R}_{n}^{(-)}(\epsilon)
  \;  .
\label{eq:gG_cond}
\end{equation}
Condition~(\ref{eq:gG_cond}),
in general, cannot be satisfied for all bound states, but it
should be satisfied, in particular, for the ground state,
so that (from Eqs.~(\ref{eq:master_eigenvalue_function_remainder_GS})
and (\ref{eq:lambda_epsilon}))
 \begin{equation}
\lambda(\epsilon)
=
\left[
\,
\Xi_{_{\rm (gs)}}  (\epsilon)
\right]^{-1}
\,
\left[ 1 +
\frac{\epsilon}{2}
\, g^{(0)}
+ o(\epsilon)  \right]
\;
\label{eq:epsilon_dependence_coupling}
\end{equation}
 and (from Eqs.~(\ref{eq:master_eigenvalue_function_remainder_GS})
and (\ref{eq:BS_energies_asymptotic3}))
 \begin{equation}
\mid \eta_{_{\rm (gs)}} \mid
= e^{  g^{(0)}  }
\;  .
\label{eq:E_renorm}
\end{equation}

Once the ground-state renormalization is established, one can
analyze the existence and properties of the excited states.
In this regard,
Eq.~(\ref{eq:critical_inequality})
selects only a subset of the states of the regularized theory
as bound states of the renormalized theory, once the limit
$\epsilon \rightarrow 0$ is taken.
More precisely, for any set of quantum numbers $n$ for which
Eq.~(\ref{eq:critical_inequality}) is a strict inequality,
the coupling becomes weak in the limit $\epsilon=0^{+}$,
so that the given state is merged with the continuum.
Thus, the equality
\begin{equation}
 \lambda^{(\ast)}
=
 \lambda^{(\ast)}_{n}
\;
\label{eq:critical_equality}
\end{equation}
is a necessary condition for the state labeled by $n$ to survive
as a bound state.
Then, for any state that satisfies Eq.~(\ref{eq:critical_equality}),
the function ${\mathcal R}_{n}(\epsilon)$ defined in
(\ref{eq:master_eigenvalue_function_remainder}) is
constrained by the limiting form
\begin{equation}
\epsilon \, {\mathcal R}_{n} (\epsilon) = o(1)
\;
\end{equation}
and (from 
Eqs.~(\ref{eq:master_eigenvalue_function})
and (\ref{eq:critical_inequality}))
satisfies the inequality
\begin{equation}
{\mathcal R}_{n} (\epsilon) \leq
 0
\;  ;
\label{eq:remainder_inequality}
\end{equation}
in particular,
\begin{equation}
{\mathcal R}_{n}^{(-)} (\epsilon) \leq
 0
\;  .
\label{eq:remainder_minus_inequality}
\end{equation}
If the inequality~(\ref{eq:remainder_minus_inequality})
were strict, then Eq.~(\ref{eq:excited_energies_asymptotic1})
would annihilate the state labeled by $n$
in the bound-state sector by exponential suppression;
as a consequence, the only alternative option for the state to
``survive,'' as allowed by the
inequality~(\ref{eq:remainder_minus_inequality}), is
\begin{equation}
{\mathcal R}_{n}^{(-)}
(\epsilon)
=
0
\label{eq:excited_energies_residue_minus}
\;  ,
\end{equation}
in which case
\begin{equation}
 \left|
 \frac{\eta_{n}(\epsilon) }{ \eta_{_{\rm (gs)}}
(\epsilon)  }
  \right|
=
\exp \left[
 {\mathcal R}_{n}^{(0)}
\right]
\;   .
\label{eq:excited_energies_asymptotic2}
\end{equation}
It should be pointed out that,
when Eqs.~(\ref{eq:critical_equality}) and
(\ref{eq:excited_energies_residue_minus})
are satisfied, the inequality
 ${\mathcal R}_{n}^{(0)}  <   0$
(from Eq.~(\ref{eq:remainder_inequality}))
guarantees that $E_{n} > E_{_{\rm (gs)}} $.

In summary, Eqs.~(\ref{eq:critical_equality}),
(\ref{eq:excited_energies_residue_minus}),
and (\ref{eq:excited_energies_asymptotic2})
give the following conditions for the existence of bound states.
An excited
state labeled with the index $n \neq {\rm (gs) } $
exists if:

(i) The critical coupling
 $ \lambda^{(\ast)}_{n}$ satisfies the
equality~(\ref{eq:critical_equality}).

(ii) The function ${\mathcal R}_{n}^{(-)} (\epsilon)  $
is identically zero
(condition~(\ref{eq:excited_energies_residue_minus}))
for the states that already satisfy
Eq.~(\ref{eq:critical_equality}).

(iii) ${\mathcal R}_{n}^{(0)}
\neq
 {\mathcal R}_{_{\rm (gs)}}^{(0)}=0$.

These are indeed very stringent conditions, so ``ordinarily''
dimensional transmutation will produce a single bound state,
as is the case for the two-dimensional delta-function and
inverse square potentials (Ref.~\cite{cam:00b}).

A final digression on strategy may provide a more
direct path in a typical problem.
If the energy generating function  $\Xi_{n}(\epsilon)$
admits the expansion
\begin{equation}
\Xi_{n}(\epsilon)
=
\left[ L_{n}  (\epsilon) \right]^{-1}
\,
\left[ 1 + \frac{\epsilon}{2}  \,
{\mathcal G}_{n} (\epsilon)   \right]
\;  ,
\label{eq:master_eigenvalue_function_practical_expansion}
\end{equation}
with a power-law leading term
\begin{equation}
L_{n}  (\epsilon)
= a_{n} \,
\epsilon^{\tau_{n}}
\;
\label{eq:master_eigenvalue_function_leading_term}
\end{equation}
(where  $ a_{n}$ and $\tau_{n}$ are constants),
and
\begin{equation}
\epsilon \,
{\mathcal G}_{n} (\epsilon) = o(1)
\;  ,
\label{eq:master_eigenvalue_G_function}
\end{equation}
then the following results will directly apply.
First, the critical coupling will be
(from Eqs.~(\ref{eq:lambda_star}),
(\ref{eq:master_eigenvalue_function_practical_expansion}),
and (\ref{eq:master_eigenvalue_G_function}))
\begin{equation}
\lambda^{(\ast)}_{n} =
\lim_{\epsilon\rightarrow 0}
L_{n}  (\epsilon)
\;  ,
\label{eq:lambda_star_practical}
\end{equation}
with a regularized coupling
(Eq.~(\ref{eq:epsilon_dependence_coupling}))
 \begin{equation}
\lambda(\epsilon)
=
L_{_{\rm (gs)}}
  (\epsilon)
\,
\left\{
 1 +
\frac{\epsilon}{2}
\,
\left[
g^{(0)}
-
{\mathcal G}_{_{\rm (gs)}}
 (\epsilon)
\right]
\right\}
+ o(\epsilon)
\;  ,
\label{eq:epsilon_dependence_coupling_practical}
\end{equation}
while the function ${\mathcal R}_{n} (\epsilon)$
of Eq.~(\ref{eq:master_eigenvalue_function_remainder})
will become
\begin{equation}
{\mathcal R}_{n} (\epsilon)
 \stackrel{\scriptscriptstyle (\epsilon \rightarrow 0)}{\sim}
{\mathcal G}_{n} (\epsilon)
-
{\mathcal G}_{_{\rm (gs)}} (\epsilon)
\;  ,
\label{eq:master_eigenvalue_function_remainder_from_G}
\end{equation}
up to higher order corrections.
Thus,
the condition for the existence of the ground state will be
$\tau_{_{\rm (gs)}} \geq 0 $, while
the conditions for the existence of excited states
will amount to the existence of
an index $n \neq {{\rm (gs)}} $,
such that: (i) $a_{n} = a_{_{\rm (gs)}}$ and
$\tau_{n} = \tau_{_{\rm (gs)}}$;
(ii) ${\mathcal G}^{(-)}_{n} (\epsilon)
=
{\mathcal G}^{(-)}_{_{\rm (gs)}}  (\epsilon)
$; and
(iii) ${\mathcal G}_{n}^{(0)}
\neq {\mathcal G}_{_{\rm (gs)}}^{(0)}
$.

We now turn to the scattering problem.

\subsection{Renormalized Scattering Sector}
\label{sec:scattering_sector_renormalized}

For the scattering sector,
the scattering amplitude $f^{(D)}_{k}(\Omega^{({D})}) $ and
the differential  scattering cross section
$d \sigma^{({D})}
(k, \Omega^{({D})})/
d \Omega_{{D}} =|f^{(D)}_{k}(\Omega^{({D})})|^{2} $
are functions of the energy $E=k^{2}$ associated
with the incident momentum $k$, as well as of the
hyperspherical angles
$\Omega^{({D})} $
(with $d \Omega_{{D}}$ being the
corresponding element of the $D$-dimensional solid angle;
see Appendix~\ref{sec:D-dim}).

As discussed in Subsection~\ref{sec:scaleinv}
and using the language developed in
Subsections~\ref{sec:BS_sector_regularized} and 
\ref{sec:BS_sector_renormalized},
there are two distinct regimes for scattering.

In the weak-coupling regime,
$\lambda < \lambda^{(\ast)}$,
the scattering is well defined for all values
of the coupling constant (consistent with the inequality defining
the weak-coupling regime).
In particular, this scattering is either
scale-invariant (energy-independent) or trivial
and needs no regularization whatsoever.
For example, the inverse square potential
gives scale-invariant scattering~\cite{mot:49,jac:72}, when
$\lambda< \left( l+D/2-1 \right)^{2}$ (where $l$ is the angular momentum
quantum number),
while the two-dimensional delta potential yields no scattering
for $\lambda<0$. These examples and issues will be analyzed in greater 
detail
in the second paper in this series.

Instead, in the strong-coupling regime,
$\lambda > \lambda^{(\ast)}$,
the coupling constant gets renormalized according to the theory 
developed
in Subsection~\ref{sec:BS_sector_renormalized}.
In particular, this implies that the coupling parameter of the
regularized theory is $\epsilon$-dependent, as displayed in
Eq.~(\ref{eq:epsilon_dependence_coupling}),
with a limiting critical value
$\lambda= \lambda^{(\ast)}+0^{+}$.
Moreover, Eq.~(\ref{eq:eigenvalues_general}) is still
applicable, as it was derived solely using scaling arguments---but
the function $\Xi(\epsilon)$ will now have a different specific form,
one that is no longer discrete.
Then, the scattering matrix and amplitude are determined from
the asymptotic
form of the scattering wave function, which is an appropriate linear
combination of scattering solutions with arguments
 $kr= \left( \widehat{\eta} \right)^{1/2}
 \; | \widehat{\mbox{\boldmath $ \xi$}} |
= \left[  \Xi(\epsilon)  \right]^{1/\epsilon}
| \widehat{\mbox{\boldmath $ \xi$}} |
 $.
Equation~(\ref{eq:eigenvalues_general}) then implies
that the scattering depends upon
\begin{equation}
| \eta (\epsilon) |^{-\epsilon/2}
= 1 -
\frac{\epsilon}{2} \, \ln \left(k^{2}/\mu^{2} \right)
\;  ,
\end{equation}
which, after taking the limit $\epsilon=0^{+}$ and
using Eq.~(\ref{eq:epsilon_dependence_coupling}), should
provide a breakdown of the scale symmetry through
the logarithmic dependence
$\ln \left(k^{2}/\mu^{2} \right)$. This suggests  that
the scattering amplitude should be of the form
\begin{equation}
f^{(D)}_{k}(\Omega^{({D})})
=
F
\mbox{\boldmath\large  $\left(  \right.$ } \! \! \!
 k,
\ln \left(k^{2}/\mu^{2} \right)   ,
\Omega^{({D})}
\mbox{\boldmath\large  $\left.  \right)$ } \! \!
\;  ,
\label{eq:scatt_amplitude_dimensional}
\end{equation}
where $F$ is a dimensional quantity. This procedure will be
illustrated for the two-dimensional delta-function potential in
Section~\ref{sec:delta}.

In Eq.~(\ref{eq:scatt_amplitude_dimensional}),
the variable $k^{2}/\mu^{2} $ explicitly appears in a
characteristic logarithmic form. However,
the function $F$ is not dimensionless
so that its form can be simplified by the $\Pi$ theorem.
We now turn to such dimensional considerations.

\subsection{Dimensional Analysis Revisited}
\label{sec:DA_renormalized}

Let us now rephrase some of the results
of Subsections~\ref{sec:BS_sector_renormalized}
and \ref{sec:scattering_sector_renormalized}
in terms of dimensional variables.

The dimensional bound-state energies are arranged
in a spectrum
\begin{equation}
E_{n}= \mu^2 \eta_{n}
\;  ,
\label{eq:GS_Pi}
\end{equation}
a conclusion that can be directly drawn from
dimensional analysis.
In particular, the ground state
defines a conventional characteristic scale
\begin{equation}
E_{_{\rm (gs)}}
= \mu^2 \eta_{_{\rm (gs)}}
\leadsto - \mu^2
\;  ,
\label{eq:GS_DA_renormalized}
\end{equation}
where the symbol
$\leadsto$ refers to the freedom to make a convenient
choice of $g^{(0)}$,
due to its arbitrariness; in this case, we selected
$g^{(0)}=0$.
This point will be further discussed and illustrated  in
Subsection~\ref{sec:delta_bound_states} for the particular case
of the two-dimensional delta-function potential.
Equation~(\ref{eq:GS_DA_renormalized})
displays in its purest form the emergence of an energy scale from the
renormalization procedure;
in addition, it
shows that naive generalized dimensional analysis
(including renormalization parameters according to
Eq.~(\ref{eq:Pi_DOF_mod}))
gives straightforwardly the correct result.

Of course, Eq.~(\ref{eq:GS_Pi})
also refers to the excited states, if they exist.
Again, by the generalized $\Pi$ theorem,
the only remaining information
about the spectrum is conveyed by the ratios
(cf.\ Eqs.~(\ref{eq:excited_energies_asymptotic2})
and (\ref{eq:master_eigenvalue_function_remainder_from_G}))
\begin{equation}
\rho_{n} =
\frac{ \eta_{n} }{ \eta_{_{\rm (gs)}} }
=
\exp
\left[
{\mathcal R}_{n} (\epsilon)
 \right]
\stackrel{\scriptscriptstyle (\epsilon \rightarrow 0)}{\sim}
\exp
\left[
{\mathcal G}_{n} (\epsilon)
-
{\mathcal G}_{_{\rm (gs)}} (\epsilon)
\right]
\;  ,
\label{eq:excited_energies_asymptotic3}
\end{equation}
 which give its characteristic ``structure,''
with the restrictions discussed in 
Subsection~\ref{sec:BS_sector_renormalized}.

For the scattering sector,
as the  dimensionality of the cross section is  $-(D-1)$
(``area'' of hypersuface),
it follows that
Eq.~(\ref{eq:scatt_amplitude_dimensional}) can be rewritten in
the form
\begin{equation}
f^{(D)}_{k}(\Omega^{({D})})
= k^{-({D} -1)/2}
\Pi
\mbox{\boldmath\large  $\left(  \right.$ } \! \! \!
\ln \left(k^{2}/\mu^{2} \right)   ,
\Omega^{({D})}
\mbox{\boldmath\large  $\left.  \right)$ } \! \!
\;  ,
\label{eq:scatt_amplitude_Pi}
\end{equation}
where $\Pi  (u, \Omega^{(D)})$ is a dimensionless function
of the dimensionless ratio $ u= (k/\mu)^{2}$.
Equation~(\ref{eq:scatt_amplitude_Pi}) will  be valid,
whether the system is capable of producing bound states or not;
in the weak-coupling regime, the function $\Pi$ is
identically constant.
On the other hand, if there is at least one bound state,
the existence of a
characteristic energy scale $E_{_{\rm (gs)}}$,
Eq.~(\ref{eq:GS_DA_renormalized}), yields an alternative form of
(\ref{eq:scatt_amplitude_Pi}),
\begin{equation}
f^{(D)}_{k}(\Omega^{({D})})
= k^{-({D} -1)/2}
\check{\Pi}
\mbox{\boldmath\large  $\left(  \right.$ } \! \! \!
\ln \left(k^{2}/ E_{_{\rm (gs)}}   \right)   ,
\Omega^{({D})}
\mbox{\boldmath\large  $\left.  \right)$ } \! \!
\;  ,
\label{eq:scatt_amplitude_Pi2}
\end{equation}
where $\check{\Pi}
\mbox{\boldmath\large  $\left(  \right.$ } \! \! \!
\ln \left(E/E_{_{\rm (gs)}}   \right)
,
\Omega^{({D})}
\mbox{\boldmath\large  $\left.  \right)$ } \! \!
=
\Pi
\mbox{\boldmath\large  $\left(  \right.$ } \! \! \!
\ln \left(E/
\mu^{2}
\right)
,
\Omega^{({D})}
\mbox{\boldmath\large  $\left.  \right)$ } \! \!
$
is another dimensionless function.
In fact, when the assignment $g^{(0)}=0$ is made
(Eq.~(\ref{eq:GS_DA_renormalized})),
the simple identity
$\check{\Pi}  =\Pi$ takes place.

Equations~(\ref{eq:GS_Pi}),
(\ref{eq:scatt_amplitude_Pi}), and (\ref{eq:scatt_amplitude_Pi2})
are just a consequence of generalized dimensional analysis.

\section{TWO-DIMENSIONAL DELTA-FUNCTION POTENTIAL}
\label{sec:delta}

One of the most basic properties of a quantum field theory is
locality, which leads to a nonrelativistic limit
with  highly singular potentials of zero range, also known as
pseudopotentials~\cite{bet:35}.
The simplest pseudopotentials are plain delta functions,
which display a large number of unusual features;
however,
in this section we will only explore those properties that
relate to the dimensional transmutation produced by the
two-dimensional representative of this class.
As we will see, this potential displays all the characteristic
fingerprints of dimensional transmutation that we discussed in
previous sections.
In fact, the two-dimensional delta-function potential
has been extensively studied in the literature,
mainly using cutoff regularization~\cite{hua:82,jac:91}
and square-well regularization~\cite{don:89}.
In our approach, we will exclusively use
dimensional regularization within the framework
defined in Sections~\ref{sec:DT_via_DR} and
\ref{sec:renormalization}.

Our strategy will be to compare the calculations with the predictions
and requirements of the general theory of
 Sections~\ref{sec:DT_via_DR} and
\ref{sec:renormalization}.
However, we will use the dimensional form~(\ref{eq:Schr1nu2}) of the
Schr\"odinger equation
from scratch, rather than any of the dimensionless
equations~(\ref{eq:Schr2nu1}) or
(\ref{eq:Schr2nu2}).
In effect, the dimensionless counterparts are most
useful at a theoretical level,
in establishing the relationships between all
relevant parameters for our problem; yet, in practice, it is
more direct to set up the ``ordinary'' (dimensional)
dimensionally regularized Schr\"odinger equation.

\subsection{Dimensional
Regularization of the Two-Dimensional Delta-Function Potential}
\label{sec:delta_DR}

The two-dimensional delta-function
potential is a particular zero-range or contact interaction of the
form
\begin{equation}
V({\bf r})= - \lambda \,
\delta^{(2)}({\bf r}) \;  .
\end{equation}
We have already seen that this potential is scale-invariant. Using
the techniques of Section~\ref{sec:DT_via_DR}, we now apply the
dimensional-continuation prescription of
Eq.~(\ref{eq:dimcontinuation}) to $W^{(2)} ({\bf r}) =
\delta^{(2)} ({\bf r})$, with $D_{0}=2$, i.e.,
\begin{eqnarray}
W^{({D})}({\bf r}_{{D}})
& = &
\int \frac{ d^{{D}} k_{{D}}}
 {(2\pi)^{  {D} }}
   \; e^{i{\bf k}_{{D}}
\cdot {\bf r}_{{D}} }
\,
\left[
\int d^{2} r_{\!2}
    \; e^{-i{\bf k}_{2} \cdot {\bf r}_{2}}
\, \delta^{(2)}({\bf r}_{2} )
\right]_{ {\bf k}_{2} \rightarrow {\bf k}_{{D}} }
\;  ,
\nonumber \\
& = &
\int \frac{ d^{{D}}
   k_{ {D}
    }} {(2\pi)^{{D}}}
   \; e^{i{\bf k}_{{D}}
\cdot {\bf r}_{{D}} }
\,
\left[ 1 \right]_{ {\bf k}_{2} \rightarrow {\bf k}_{{D}} }
\nonumber \\
& = &
\int \frac{ d^{{D}} k_{{D} }}
{(2\pi)^{{D}}}
   \; e^{i{\bf k}_{{D}}
\cdot {\bf r}_{{D}} }
=
\delta^{({D})} ({\bf r})
\;  ,
\label{eq:dimcontinuation_delta}
\end{eqnarray}
which is the obvious dimensional extension of the
original delta-function potential.
Thus, in what follows,
we will consider the dimensionally regularized problem
\begin{equation}
[- \nabla_{{\bf r},
{D} }^2   -
    \lambda \, \mu^{\epsilon} \,
\delta^{({D})}
 ( {\bf r} )  ]
\,  \Psi (  {\bf r} )
= E
 \, \Psi (  {\bf r} )
\;  .
\label{eq:Schr2nu1delta}
\end{equation}

Straightforward
solution of Eq.~(\ref{eq:Schr2nu1delta}) in this context
should {\em not\/} be interpreted as a way
of drawing conclusions about the $D$-dimensional delta-function
potential. Instead, it is just the means to regularize
the $D_{0}=2$ case.
Of course, one could adjust the regularization
to be applied around a value $D_{0} \neq 2$;
however, we will not attempt such modification in this paper, as it
is not directly relevant to dimensional transmutation.

\subsection{Bound-State Sector for a Two-Dimensional
Delta-Function Potential}
\label{sec:delta_bound_states}

Equation~(\ref{eq:Schr2nu1delta})
can be easily solved in momentum space;
for the bound-state sector,
\begin{equation}
\widetilde{\Psi}({\bf q}) =
\lambda \, \mu^{\epsilon}
\,
\frac{\Psi ({\bf 0})}{q^{2} - E}
\;  ,
\label{eq:delta_reciprocal_wf}
\end{equation}
which, via the inverse Fourier
transform, yields the position-space eigenfunctions. However,
if we are only interested in the eigenvalue equation,
it suffices to consider the value of the wave function at the origin,
\begin{equation}
\Psi ({\bf 0})
=
\int
\frac{d^{{D}}q}{ (2\pi)^{{D}}}
\,
\widetilde{\Psi}({\bf q})
\;   ,
\end{equation}
so that Eq.~(\ref{eq:delta_reciprocal_wf}) gives
\begin{equation}
\frac{\lambda \, \mu^{\epsilon}}{(2\pi)^{{D}}}
\int
\frac{d^{{D}}q}{q^{2}+|E|} = 1
\;  ,
\label{eq:delta_eigen_Fourier}
\end{equation}
where $E= - | E |$
($E<0$ for the possible bound states).
Equation~(\ref{eq:delta_eigen_Fourier})
can be straightforwardly integrated using
Eq.~(\ref{eq:beta_D_integral}), which implies that
\begin{equation}
\int
\frac{d^{{D}}q}{q^{2}+|E|} =
\pi^{{D}/2}
|E|^{{D}/2 - 1}
\Gamma \left( 1 - \frac{D}{2} \right)
\; ,
\label{eq:delta_eigen_Fourier_aux}
\end{equation}
and the regularized eigenvalue equation takes the form
\begin{equation}
\frac{\lambda \, \mu^{\epsilon}}{4\pi}
\left(
\frac{|E|}{4\pi}
\right)^{ {D} /2 - 1}
\Gamma \left(1 - \frac{D}{2} \right) =
 1
\; .
\label{eq:delta_eigen}
 \end{equation}
It is a simple exercise to verify that
Eq.~(\ref{eq:delta_eigen}) reduces to the familiar textbook result
$\kappa=\sqrt{|E|}=\lambda/2$
for $D=1$~\cite{rob:97}.
On the other hand,
the left-hand side is divergent for $D=$ $2,4,6, \ldots$.
However, the restriction on the spatial dimension $D$ of regular 
potentials
is even stronger because more stringent conditions
are dictated by the eigenfunctions, as we will see next.
In our subsequent analysis, both for two-dimensional
delta-function and inverse square potentials,
the dimension $D$ will usually
appear in terms of the variable
\begin{equation}
\nu= D/2 -1
\; ,
\label{eq:nu_def}
\end{equation}
which will thereby simplify the form of most formulas; for example,
the eigenvalue Eq.~(\ref{eq:delta_eigen})
reads (with $\epsilon=-2 \nu$)
\begin{equation}
\frac{\lambda \, \mu^{-2\nu}}{4\pi}
\left(
\frac{|E|}{4\pi}
\right)^{\nu}
\Gamma
\left( -\nu \right) = 1
\; .
\label{delta_eigen3}
\end{equation}
Then the inverse Fourier transform,
\begin{equation}
\Psi( {\bf r} )
=
\lambda \, \mu^{\epsilon}
\,
\Psi({\bf 0})
\int
\frac{e^{i{\bf q}\cdot
{ {\bf r}}}}
{q^{2}+|E|}
\frac{d^{{D}}q}{(2\pi)^{{D}}}
\; ,
\label{eq:Fourier_inverse}
\end{equation}
is recognized to be proportional to the Green's function
${\mathcal K}_{{D}}({\bf r}; \kappa)$
for the modified Helmholtz equation
(see Appendix~\ref{sec:D-dim_GF}, in particular
Eqs.~(\ref{eq:Helmholtz_mod})--(\ref{eq:DD_GF2_mod})),
\begin{eqnarray}
\Psi( {\bf r} )  & = &
- \lambda \, \mu^{\epsilon}
\Psi({\bf  0})
\,
{\mathcal K}_{{D}}({\bf r}; \kappa)
\label{eq:delta_wf_green}
\\
& = &
\frac{  \lambda \, \mu^{\epsilon}  \Psi({\bf  0}) }{ 2 \pi }
\left(
\frac{\kappa}{2 \pi r}
\right)^{ \nu}K_{ \nu }(\kappa r)
\; ,
\label{eq:delta_wf}
\end{eqnarray}
where $\kappa= \sqrt{|E|}$,
$r=|{\bf r} |$, and
 $K_{\nu}(z)$ is the modified Bessel function  of the second
kind of order $\nu$.
The asymptotic behavior of the wave function $\Psi({\bf r})$
(and of the Green's function
${\mathcal K}_{{D}}({\bf r}; \kappa)$)
 is governed by that of  $K_{\nu}(z)$~\cite{gra:80},
\begin{equation}
K_{\nu}(z)
 \stackrel{\scriptscriptstyle (z \rightarrow \infty)}{\sim}
\sqrt{\frac{\pi}{2z}} \, e^{-z} \left[ 1 + O(1/z) \right]
\;  ,
\end{equation}
whence
\begin{equation}
\Psi({\bf r})
 \stackrel{\scriptscriptstyle (r \rightarrow \infty)}{\sim}
 \frac{\lambda \, \mu^{\epsilon} \,  \Psi ({\bf 0})}{4\pi}
\left( \frac{\kappa}{2 \pi} \right)^{({D}-3)/2}
\,
\frac{e^{-\kappa r}}{r^{( {D} -1)/2}}
\left[ 1 + O(1/r) \right]
\; ,
\label{eq:delta_wf_asymptotic}
\end{equation}
which
displays the correct  behavior
for a bound-state wave function at infinity.
However, near the origin,
the modified Bessel function has a singular
 behavior~\cite{abr:72}
of the form
\begin{eqnarray}
  K_{p} (z)
&  \stackrel{(z \rightarrow 0)}{\sim} &
 \frac{1}{2}
\,
\left[
\Gamma (p)
\left( \frac{z}{2} \right)^{-p}
 +
\Gamma (-p)
\left( \frac{z}{2} \right)^{p}
\right]
\,
\left[ 1 + O(z^{2}) \right]
\;  ,
\\
&  \stackrel{(z \rightarrow 0)}{\sim} &
\left\{ \begin{array} {ll}
\frac{1}{2} \,
\Gamma (|p|) (2/z)^{|p|}
\,
\left[ 1 + O(z^{2}) \right]
  &  {\rm for} \; p \neq 0
\\   - \left[ \ln (z/2) + \gamma \right]
\,
\left[ 1 + O(p,z^{2}) \right]
 & {\rm for} \;   p \approx 0
 \end{array}
\right.
\;  ,
\label{mod_Bessel_small_arg}
\end{eqnarray}
where $\gamma$ is the Euler-Mascheroni constant;
then, the explicit form of the wave function
is
\begin{equation}
\Psi({\bf r})
 \stackrel{\scriptscriptstyle (r \rightarrow 0)}{\sim}
 \frac {\lambda \, \mu^{\epsilon} \Psi ({\bf 0})}{4\pi}
\left(
\frac{\kappa^{2}}{4\pi}  \right)^{ \nu}
\times
\left\{ \begin{array} {ll}
 \Gamma (\nu) \,
(\kappa
r/2)^{-2\nu}  &  {\rm for} \; D>2
\\ -2  \left[ \ln( \kappa r/2) + \gamma   \right]  & {\rm for} \; D  =
2 \\
\Gamma(-\nu) &  {\rm for}
\; 0<D<2 \end{array}
\right.
\;  ,
\label{eq:delta_wf_origin}
\end{equation}
which shows that the nature of the solution changes around
$\nu=0$, i.e., for $D=2$.
This confirms the {\em critical\/} character of the
dimension $D=2$ for the delta-function potential.
Notice that the wave functions are singular
at the origin for any dimension $D \geq 2$.
Parenthetically, this is an example of an ultraviolet divergence:
the wave function is singular at small distances
or due to large momenta (cf.\ Eq.~(\ref{eq:delta_eigen_Fourier})).
For $D<2$, we can regularize the two-dimensional delta-function
potential and take the limit $\nu \rightarrow 0$ in
Eq.~(\ref{eq:delta_wf_origin}),
thus recovering self-consistently
the eigenvalue equation~(\ref{delta_eigen3}),
which, with $D=2-\epsilon$, i.e., $\nu=-\epsilon/2$,
reads
\begin{equation}
\frac{\lambda \, \mu^{\epsilon}}{4\pi}
\left(
\frac{|E|}{4\pi}
\right)^{-\epsilon/2}
\Gamma
\left(\frac{\epsilon}{2} \right) = 1
\; ,
\label{delta_eigen2}
\end{equation}
Alternatively,
in the language of Eq.~(\ref{eq:delta_wf}), this
eigenvalue equation can be simply enforced by the condition
\begin{equation}
 - \lambda \, \mu^{\epsilon} {\mathcal K}_{{D}} ({\bf 0}; \kappa)
= 1
\;  .
\label{eq:delta_eigen_green}
\end{equation}

Having completed the exploratory analysis of the bound-state
sector and found the eigenvalue equation
((\ref{eq:delta_eigen}), (\ref{delta_eigen3}),
 (\ref{delta_eigen2}), or (\ref{eq:delta_eigen_green})),
we are ready to compare these expressions with the
general eigenvalue equation~(\ref{eq:eigenvalues_BS}),
which will now include an energy generating function
\begin{equation}
\Xi (\epsilon) =
\frac{1}{4\pi} \left(4\pi\right)^{\epsilon/2 }
\Gamma
\left(\frac{\epsilon}{2}
\right)
\; .
\label{eq:delta_master_eigenvalue_function}
\end{equation}
It is immediately
apparent from Eq.~(\ref{eq:delta_master_eigenvalue_function})
that the theory has only one bound state, so that there is no need
for a quantum number.
In order to determine whether this ground state survives
the renormalization process,
we should look at the $\epsilon=0^{+}$ limit
of Eq.~(\ref{eq:delta_master_eigenvalue_function}),
which can be conveniently examined through the expansion
\begin{equation}
\Xi (\epsilon) = \frac{1}{2\pi\epsilon}
\left[  1 +
\frac{\epsilon}{2}
\,
\left( \ln 4\pi - \gamma \right)
+ O(\epsilon^{2}) \right]
\;  .
\label{eq:delta_master_eigenvalue_function_expansion}
\end{equation}
From Eqs.~(\ref{eq:lambda_star}) and
(\ref{eq:delta_master_eigenvalue_function_expansion}),
 the critical coupling is found to be
\begin{equation}
\lambda^{(\ast)}=0
\; ,
\end{equation}
so that the theory looks asymptotically free but still
engenders a unique bound state.
It should be noticed that this is achieved by
the regularization of the coupling constant
through the strategy of Eq.~(\ref{eq:epsilon_dependence_coupling})
(or Eq.~(\ref{eq:epsilon_dependence_coupling_practical})),
so that
\begin{equation}
\lambda(\epsilon) = 2 \pi \epsilon
\left\{ 1 + \frac{\epsilon}{2}  \,
\left[
g^{(0)}
- \left(  \ln 4\pi - \gamma \right)
\right]
\right\}
\; ,
\label{eq:delta_renormalized_coupling}
\end{equation}
leading to a ground state
\begin{equation}
E_{_{\rm (gs)}}
=
- \mu^{2}
e^{ g^{(0)}  }
\;  .
\label{delta_gs}
\end{equation}

A final remark about renormalization shows
additional parallels with the
corresponding field-theory problems.
As usual, the arbitrariness in the choice
of the finite part $g^{(0)}  $ can be used to simplify the expressions 
above
in such a way that $|E_{_{\rm (gs)}}|=\mu^{2}$,
as displayed in Eq.~(\ref{eq:GS_DA_renormalized}).
On the other hand, the singular nature of the ground state has been 
tamed by
subtracting the divergent part of
Eq.~(\ref{eq:delta_master_eigenvalue_function}),
which amounts to the subtraction of the pole $1/2\pi \epsilon$.
However, due to the arbitrariness in the choice of
$g^{(0)}$, at the level of the ground-state energy,
we have also subtracted---along with the pole---the term
$  \ln 4\pi - \gamma$ (which is an artifact of the
dimensional-regularization technique).
This is  recognized to be the
usual modified minimal subtraction ($\overline{\rm MS}$)
scheme~\cite{pes:95}.

In conclusion, the unregularized problem
has a singular spectrum
with a unique energy level at $-\infty$ and vanishing critical
coupling.
The  regularization process brings this level
to a finite value, which, upon renormalization,
becomes the  unique ground state of the
two-dimensional delta-function potential.

\subsection{Scattering Sector for a Two-Dimensional Delta-Function 
Potential}
\label{sec:delta_scattering}

The scattering sector of the Schr\"odinger equation
is described by its equivalent Lippmann-Schwinger
equation~(\ref{eq:LS}), which,
for a two-dimensional delta-function potential
(\ref{eq:Schr2nu1delta}),
takes the simple form
\begin{equation}
\Psi^{(+)} ({\bf r})
= e^{i {\bf k} \cdot {\bf r}}
- \lambda \, \mu^{\epsilon}
\,
{\mathcal G}^{(+)}_{{D}} ({\bf r}; k)
     \Psi^{(+)} ({\bf 0})
\; ,
\label{eq:LS_delta}
\end{equation}
where ${\mathcal G}^{(+)}_{{D}} ({\bf r}; k)$
is the Green's function for the Helmholtz equation,
with outgoing boundary conditions
(see Appendix~\ref{sec:D-dim_GF}, in particular
Eqs.(\ref{eq:Helmholtz})--(\ref{eq:DD_GF2})).
In particular, Eq.~(\ref{eq:LS_delta}) implies that
\begin{equation}
    \Psi^{(+)} ({\bf 0})
=
\left[ 1 + \lambda \, \mu^{\epsilon}
\,
{\mathcal G}^{(+)}_{{D}} ({\bf 0};k) \right]^{-1}
\;  .
\label{eq:Psi_0_delta}
\end{equation}
The asymptotic form of Eq.~(\ref{eq:LS_delta})
is obtained as described in Appendix~\ref{sec:D-dim_scattering},
according to which (e.g., Eqs.~(\ref{eq:DD_GF_asymptotic})
and (\ref{eq:DD_LS_asymptotic})) the scattering amplitude becomes
\begin{equation}
f^{(D)}_{k}
( \Omega^{({D})}   )
=
- \Gamma_{{D}} (k)
\lambda \, \mu^{\epsilon} \,
    \Psi^{(+)} ({\bf 0})
\; ,
\label{eq:f_delta}
\end{equation}
where
\begin{equation}
 \Gamma_{{D}} (k) =
- \frac{1}{4 \pi} \left( \frac{k}{2 \pi}
\right)^{( {D} -3)/2}
\;  .
\label{eq:Gamma_D}
\end{equation}
Finally, Eqs.~(\ref{eq:Psi_0_delta}) and
(\ref{eq:f_delta}) provide the desired expression,
\begin{equation}
f^{(D)}_{k}
( \Omega^{({D})}   )
=
- \Gamma_{{D}} (k)
\left[ \left( \lambda \, \mu^{\epsilon} \right)^{-1} +
{\mathcal G}^{(+)}_{{D}} ({\bf 0};k) \right]^{-1}
\; .
\label{eq:f_delta2}
\end{equation}

Equation~(\ref{eq:f_delta2})
is singular for $D \geq 2 $,
as can be seen from the divergent small-argument limit of
Eq.~(\ref{eq:DD_GF2}).
However, one can use the renormalization of the bound-state sector
to eliminate this divergence through the regularized coupling,
Eq.~({\ref{eq:delta_renormalized_coupling}).
More precisely,
for an attractive potential,
we found that the
coupling constant can be traded in favor of  dimensional parameters,
e.g., using the Green's function
${\mathcal K}_{{D}} ({\bf r}; \kappa)$ for the
bound-state sector in
Eq.~(\ref{eq:delta_eigen_green}).
In other words,
using the renormalization for the bound-state
sector, we will now obtain directly the renormalized scattering
amplitude, which is explicitly given by the limit
\begin{equation}
f_{k}
( \Omega^{({D})}   )
=
 \Gamma_{{D}} (k)
\lim_{r \rightarrow 0}
\left[  {\mathcal K}_{{D}} ({\bf r};
 \sqrt{E_{_{\rm (gs)}}}) -
{\mathcal G}^{(+)}_{{D}} ({\bf r}; k) \right]^{-1}
\; ,
\label{delta_scatt_ampl_diff}
\end{equation}
where the replacement
\begin{equation}
\kappa = \sqrt{|E_{_{\rm (gs)}}|}
\;
\end{equation}
was made.
Equation~(\ref{delta_scatt_ampl_diff}) already displays a remarkable
property of the scattering by a delta-function potential:
it is {\em isotropic\/},
i.e., it only scatters s-waves, as it corresponds intuitively
to a contact interaction.

Let us now evaluate the limit in Eq.~(\ref{delta_scatt_ampl_diff}).
First, from  Eqs.~(\ref{eq:delta_eigen_Fourier_aux})
and (\ref{eq:DD_GF_mod}),
\begin{equation}
\lim_{r \rightarrow 0}
  {\mathcal K}_{{D}} ({\bf r};
\sqrt{|E_{_{\rm (gs)}}|})
=
-
\frac{1}{ (2 \pi)^{D}}
\int
\frac{d^{{D}}q}{ q^{2}+  |E_{_{\rm (gs)}}| }
=
-\frac{1}{ ( 4 \pi)^{D/2} } \,
|E_{_{\rm (gs)}}|^{D/2 - 1}
\Gamma \left( 1 - \frac{D}{2} \right)
\; .
\label{eq:delta_eigen_Fourier_aux_scatt}
\end{equation}
Next,
$\lim_{r \rightarrow 0}
  {\mathcal G}^{(+)}_{{D}} ({\bf r}; k)$
can be obtained by analytic continuation
\begin{equation}
  {\mathcal G}^{(+)}_{{D}} ({\bf r}; k)
=
 \left.
{\mathcal K}_{{D}} ({\bf r}; \kappa)
\right|_{ \kappa^{2} \rightarrow - (k^{2}+ i \delta) }
\;  ,
 \end{equation}
where $\delta=0^{+}$,
which implies that
\begin{equation}
\lim_{r \rightarrow 0}
  {\mathcal G}^{(+)}_{{D}} ({\bf r}; k)
=
\frac{1}{ (2 \pi)^{D}}
\,
\int
\frac{d^{{D}}q}{k^{2}- q^{2}+i \delta}
=
-
\frac{1}{ ( 4 \pi)^{D/2} } \,
\left(-k^{2}-i \delta \right)^{D/2 - 1}
\Gamma \left( 1 - \frac{D}{2} \right)
\; .
\label{eq:delta_eigen_Fourier_aux_scatt2}
\end{equation}
Thus,
\begin{equation}
\lim_{r \rightarrow 0}
 \left[
{\mathcal G}^{(+)}_{{D}} ({\bf r}; k)
-
  {\mathcal K}_{{D}} ({\bf r}; \kappa)
\right]
=
\frac{1}{ ( 4 \pi)^{D/2} } \,
\Gamma \left( 1 - \frac{D}{2} \right)
\left[
|E_{_{\rm (gs)}}|^{ {D}/2 - 1}
- \left( -k^{2} - i \delta \right)^{ {D}/2 - 1}
\right]
\;  ,
\end{equation}
which can be evaluated in the limit
$\epsilon \rightarrow 0^{+}$, with $D=2-\epsilon$,
\begin{equation}
\lim_{\epsilon \rightarrow 0}
\lim_{r \rightarrow 0}
 \left[
{\mathcal G}^{(+)}_{{D}} ({\bf r}; k)
-
  {\mathcal K}_{{D}} ({\bf r}; \kappa)
\right]
=
-
\frac{1}{4 \pi}
\left( \ln |E_{_{\rm (gs)}}|
- \ln k^{2} + i \pi \right)
\;  ,
\label{eq:green_limit}
\end{equation}
where the identity
$\ln [- (k^{2}+ i \delta) ]=
\ln k^{2} - i\pi$ was used.
Finally, the scattering amplitude is
obtained by replacing Eqs.~(\ref{eq:Gamma_D}) (with $D=2$) and
(\ref{eq:green_limit}) in (\ref{delta_scatt_ampl_diff}), i.e.,
\begin{equation}
f^{(2)}_{k}
( \Omega^{(2)}   )
=
\sqrt{\frac{2\pi}{ k}} \,
\left[ \ln \left( \frac{k^{2}}{E_{_{\rm (gs)}}} \right)
-i \pi
\right]^{-1}
\;  .
\label{eq:delta_scattering_amplitude}
\end{equation}
Equation~(\ref{eq:delta_scattering_amplitude}) is seen to
agree with the prediction of generalized dimensional analysis,
Eqs.~(\ref{eq:scatt_amplitude_Pi})--(\ref{eq:scatt_amplitude_Pi2}),
with a dimensionless
variable $\Pi (u)=
\sqrt{2 \pi} \left[ \ln u - i \pi \right]^{-1}$, and
$u=k^{2}/E_{_{\rm (gs)}}$.

Finally, the
differential scattering cross section
$
d \sigma^{(2)}
(k, \Omega^{(2)} )/
d \Omega_{2}
$, from Eq.~(\ref{eq:diff_cross_section}),
 again agrees with the prediction of generalized dimensional analysis,
Eqs.~(\ref{eq:scatt_amplitude_Pi})--(\ref{eq:scatt_amplitude_Pi2}),
providing a dimensionless variable $\Pi (u)=
2 \pi \left[ (\ln u)^{2}+\pi^{2}\right]^{-1}$,
 for the energy ratio $u=k^{2}/E_{_{\rm (gs)}}$.

\section{CONCLUSIONS}
\label{sec:conclusions}

Until recently, it had been generally assumed that generic
field-theoretic tools and concepts are useful only for systems
with infinitely many degrees of freedom. While this perception is
essentially correct for ``regular'' systems, it is now recognized,
as discussed in our series of papers, that such techniques can be
generalized and used to extract meaningful physical results for
certain ``singular'' systems with a finite number of degrees of
freedom.

In this paper, we developed systematic uses of the techniques of
dimensional regularization and renormalization, and of the concept
of dimensional transmutation, with the purpose of gathering
information about the class of scale-invariant potentials. Our
discussion relied on dimensional regularization, which we argued
provides a generic tool for the treatment of all members of
that class, by establishing a simple link between the two meanings
of the word dimension. From our fairly general investigation, we
have learned that all scale-invariant potentials are homogeneous
of degree $-2$ and share a number of remarkable properties; here,
without attempting to give an exhaustive list, we summarize a few
of the most outstanding:

 (i) There exists a critical coupling $\lambda^{(\ast)}$ such
that, for $\lambda < \lambda^{(\ast)}$ (weak coupling) the
Hamiltonian is self-adjoint but produces no bound states, while
for $\lambda > \lambda^{(\ast)}$ (strong coupling) it loses its
self-adjoint character, generating a continuum of bound states not
bounded from below and requiring renormalization.

(ii) Solution of the regularized theory for strong coupling yields
a master eigenvalue equation~(\ref{eq:eigenvalues_general}), which
condenses all the information about the given scale-invariant
potential and requires proper renormalization.

(iii) The ground state of a given ``strong'' scale-invariant
potential exists provided that $\lambda^{(\ast)}
 = \left[
\lim_{\epsilon\rightarrow 0}
\,
\Xi_{n}(\epsilon)
\right]^{-1}$
(Eq.~(\ref{eq:lambda_star})) be finite.

(iv) Excited states exist under the demanding conditions listed
after Eq.~(\ref{eq:excited_energies_asymptotic2}). Thus,
``strong'' scale-invariant potentials have a tendency to suppress
excited states.

(v) The scattering sector remains scale-invariant or trivial in
the weak-coupling regime, while it displays a logarithmic
dependence $\ln \left( k^{2}/\mu^{2} \right)$, with respect to the
energy $k^{2}$, in the strong-coupling regime.

(vi) In short, in the strong-coupling regime, a given
scale-invariant potential leads to dimensional transmutation,
which manifests itself on the existence of at least one bound
state and a scale-dependent scattering matrix. The dimensional
transmutation exhibited for strong coupling amounts to the
emergence of a scale anomaly, i.e., quantum-mechanical breaking of
the classical scale symmetry.

Additional progress in understanding these singular
quantum-mechanical systems can best be achieved by studying
specific cases. A first attempt was made in this paper by
considering some aspects of the two-dimensional delta-function
potential. In that regard, the second paper in this
series~\cite{cam:00b} will present a more thorough analysis of the
two-dimensional delta-function potential, as well as a novel
treatment of the anomalous transmuting behavior of the inverse
square potential.

\appendix

\section{DIMENSIONAL REGULARIZATION IN
$D$-DIMENSIONAL EUCLIDEAN SPACES}
\label{sec:D-dim}

Just like for the corresponding case in quantum field 
theory~\cite{pes:95},
our approach is based on the dimensional extension of
mathematical expressions
for a system that is assumed to be embedded
in a $D$-dimensional Euclidean space.
Then,
starting from the  Cartesian coordinates
$(x_{1}, \ldots , x_{{D}})$,
one can introduce an alternative  set of $D$-dimensional
hyperspherical  polar coordinates
$( q_{{0}}=r,q_{1}=\theta_{1},
\ldots, q_{{D}-1}
=\theta_{{D}-1})$
via the transformation equations
\begin{eqnarray}
 x_{1} & = & r \cos \theta_{1} 
\nonumber \\
 x_{2} & = & r \sin \theta_{1} \cos \theta_{2} 
\nonumber \\
& \vdots  & 
\nonumber \\
x_{j} & = & r
\left( \prod_{k=1}^{j-1} \sin\theta_{k} \right) \cos \theta_{j} 
\nonumber \\
& \vdots & 
\nonumber \\
x_{{D}} & = & r
\prod_{k=1}^{{D}-1} \sin\theta_{k}
\;  ,
\end{eqnarray}
where the ranges for the hyperspherical polar variables are
$0 \leq \theta_{j} \leq  \pi$
for $j=1, \ldots ,D-2$;
$0 \leq \phi \equiv \theta_{{D}-1} \leq 2 \pi$; and
$0 \leq r < \infty$.
All the basic geometric quantities associated with
 hyperspherical coordinates can be constructed through the
scale coefficients $h_{j}$ for the diagonal metric
$(g_{ij} ) = {\rm diag} (h_{j}^{2})_{0\leq j \leq 
D-1}$~\cite{mor:53b};
these coefficients are straightforwardly given by
$h_{0}= 1$,
$h_{1}= r$,
and
$h_{j}= r   \prod_{k=1}^{j-1} \sin\theta_{k} $
(for $2 \leq j \leq D-1$),
while their product is
\begin{equation}
h^{({D})} = \prod^{{D-1}}_{j=0} h_{j} =
r^{{D}-1}
\prod^{{D}-1}_{j=1}
\sin^{{D}-j-1 }\theta_{j}
\; .
\end{equation}

In our series of papers,
both the Laplacian operator and the element of volume
are needed.
The Laplacian
can be computed from~\cite{mor:53c}
\begin{eqnarray}
\nabla^{2}_{{D}}
& =  &
\frac{1}{ h^{({D})} }
\sum^{{D-1}}_{j=0}\frac{\partial}
{\partial  q_{j}}
\left( \frac{h^{({D})}}{h_{j}^{2}}
\frac{\partial}{\partial q_{j}}    \right)
\nonumber  \\
& =  &
\Delta^{({D})}_{r}+
\frac{1}{r^{2}}
\Delta_{\Omega^{({D})}}
\;  ,
\end{eqnarray}
where its  radial part is, explicitly,
\begin{equation}
\Delta^{({D})}_{r}
=\frac{1}{ r^{{D}-1} }
\frac{\partial}{\partial r}
\left( r^{{D}-1}
\frac{\partial}{\partial r} \right)
\;  ,
\label{eq:radial_Laplacian}
\end{equation}
while its  angular part is
\begin{equation}
\Delta_{\Omega^{({D})}}
=
\sum^{{D}-1}_{j=1}
\left[
\left( \prod_{k=1}^{j-1} \sin^{2} \theta_{k} \right)
\, \sin^{{D}-j-1} \theta_{j}
\right]^{-1}
\frac{\partial}
{\partial\theta_{j}}
\left(
\sin^{{D}-j-1} \theta_{j}  \,
\frac{\partial}{\partial\theta_{j}}
\right)
\; ,
\label{eq:angular_Laplacian}
\end{equation}
in which the notation $ \Omega^{({D})}
 \equiv (\theta_{1} , \ldots , \theta_{{D}-1})$
has been introduced and it is implied
that $ \prod_{k=k_{1}}^{k_{2}}  \equiv 1$ when
$k_{1} >k_{2}$ (i.e., for $j=1$).

 Similarly, the
 element of the $D$-dimensional solid angle
becomes
\begin{equation}
d \Omega_{{D}}
=
\prod^{ {D}-1}_{j=1} h_{j} d \theta_{j}
=
h^{({D})}
\prod^{{D}-1}_{j=1} d \theta_{j}
= \prod^{{D}-1}_{j=1}
 \sin^{{D}-j-1} \theta_{j} \, d \theta_{j}
\; ,
\label{eq:diff_solid_angle}
\end{equation}
in terms of which
the $D$-dimensional volume element is given by
\begin{equation}
d^{{D}} r = r^{{D}-1}
d \Omega_{{D}} \, dr
\;  .
\label{eq:volume_element}
\end{equation}
Equation~(\ref{eq:diff_solid_angle}) can be integrated to a total
$D$-dimensional solid angle
\begin{equation}
\Omega_{{D}}=
\int d\Omega_{{D}}
=
\left(
\prod^{{D}-2}_{j=1}
\int_{0}^{\pi}  d \theta_{j}
\sin^{{D}-j-1} \theta_{j}
\right)
\int_{0}^{2\pi}  d \theta_{{D}-1}
\; ,
\end{equation}
where the angular integrals can be evaluated
using the  beta-function identity~\cite{gra:80},
\begin{equation}
\int_{0}^{\pi/2} \sin^{m} \theta   \,  d \theta
=
\frac{1}{2} B
\mbox{\boldmath\large  $\left(  \right.$ } \! \! \!
 (m+1)/2, 1/2
\mbox{\boldmath\large  $\left.  \right)$ } \! \!
= \frac{
\sqrt{\pi} \, \Gamma \mbox{\boldmath\large  $\left(  \right.$ } \! \! \!
 (m+1)/2
\mbox{\boldmath\large  $\left.  \right)$ } \! \!
}
{ 2  \;
\Gamma \mbox{\boldmath\large  $\left(  \right.$ } \! \! \!
 (m+2)/2
\mbox{\boldmath\large  $\left.  \right)$ } \! \!
}
\;  ,
\end{equation}
whence
\begin{equation}
\Omega_{{D}}
 =
\frac{2 \;\pi^{{D}
/2}}{\Gamma(D/2)}
\;  .
\end{equation}
With the given element of volume, Eq.~(\ref{eq:volume_element}),
one can compute
the integral of any function $f(r)$  that
exhibits $D$-dimensional central symmetry,
that is,
\begin{equation}
\int f(r)d^{{D}}r =
\frac{2 \;  \pi^{{D}
/2}}{\Gamma(D/2)}
\int_{0}^{\infty}
r^{{D}-1}
f(r)dr
\; .
\label{eq:central}
\end{equation}
In particular,
Eqs.~(\ref{eq:diff_solid_angle})--(\ref{eq:central})
are essential for the evaluation of expressions
in dimensional regularization,
in conjunction with another beta-function identity~\cite{gra:80},
\begin{equation}
\int_{0}^{\infty}
\frac{x^{2 \alpha-1}}{(x^{2}+1)^{\alpha+\beta}}d x =
\frac{1}{2}
B(\alpha,\beta)
\; ,
\end{equation}
whence
\begin{equation}
\int\frac{(q^{2})^{n}}{( q^{2} + a^{2} )^{m}}d^{{D}}q
 =
\pi^{D/2}
a^{D+ 2n - 2m}
\,
\frac{\Gamma(n+D/2) \, \Gamma(m - n - D/2)}{\Gamma (D/2) \, \Gamma(m)}
\;  .
\label{eq:beta_D_integral}
\end{equation}

Finally, let us consider the
 general $D$-dimensional Fourier transform $\widetilde{f}({\bf s})$
of a function $f({\bf u})$, defined by
\begin{equation}
\widetilde{f}({\bf s}) =
\frac{n_{D}}{(2\pi)^{{D}/2} }
\int d^{{D}}  u  \,
e^{\mp i{\bf s}\cdot{\bf u}}f({\bf u})
\; ,
\end{equation}
with an arbitrary normalization constant $n_{D}$
(usually $n_{D}= 1$
or $n_{D}=(2 \pi)^{\pm D/2}$).
Its computation can be simplified considerably
for the particular case when
the function displays central symmetry, i.e.,
 $f({\bf u})= f(u)$.
 In effect, for the integration of $f(u)$,
selecting coordinates according to
 ${\bf s} \cdot {\bf u} = su \cos \theta_{1}$,
it follows that
\begin{eqnarray}
\widetilde{f}({\bf s}) & = &
\frac{n_{D}}{(2\pi)^{{D}/2}}
\left(
\prod_{j=2}^{{D}-2}
\int_{0}^{\pi}   d\theta_{j}
\sin^{{D} - j-1} \theta_{j}
\right) \,
\int_{0}^{2 \pi} d\theta_{{D} - 1}
\nonumber \\
& \times &
\int^{\infty}_{0} du \,
u^{{D} - 1}
f(u) \int^{\pi}_{0} d\theta_{1}
\sin^{{D} - 2}\theta_{1}
e^{\mp is u \cos\theta_{1}}
\nonumber \\
 & = &
n_{D} (2\pi)^{-{D}/2} \,
\Omega_{{D}-1}
\int^{\infty}_{0} du  \,
u^{{D} - 1}f(u)
\, {\mathcal I} (D/2 -1, s u)
\;  ,
\end{eqnarray}
where~\cite{gra:80}
\begin{equation}
{\mathcal I} (\nu,z)
=
\int^{\pi}_{0}e^{\mp
i z\cos\theta}sin^{2 \nu}\theta d\theta =
\frac{\Gamma(\nu +1/2) \, \Gamma(1/2)}{ (z/2)^{\nu}}
\, J_{\nu}(z)
\;  ,
\end{equation}
which  implies that ($\nu=D/2-1$)
\begin{equation}
\widetilde{f}({\bf s}) =
\frac{n_{D}}{(2\pi)^{{D}/2} }
\int d^{{D}}u e^{-i{\bf s}\cdot {\bf u}} f({\bf u})
= \frac{n_{D}}{
s^{{D}/2 - 1}} \, \int^{\infty}_{0}
f(u) J_{{D}/2 -1} (su)
u^{{D}/2} \, du
\;  .
\label{eq:Bochner}
\end{equation}
Equation~(\ref{eq:Bochner}),
which is  a Hankel transform,
 is sometimes referred to as Bochner's theorem.

\section{$D$-DIMENSIONAL GREEN'S FUNCTIONS}
\label{sec:D-dim_GF}

As an example of Bochner's theorem, we will now compute the
infinite-space Green's function
 for the $D$-dimensional Helmholtz equation.
We will start with the modified
 Helmholtz equation,
\begin{equation}
\left[
 \nabla_{{\bf  r},{D}}^{2} -
\kappa^{2}
  \right]
K_{{D}} ({\bf  r},{\bf  r}'; \kappa)
=
\delta^{({D})}
({\bf  r} - {\bf  r}')
\; ,
\label{eq:Helmholtz_mod}
\end{equation}
whose Green's function
$K_{{D}} ({\bf  r},{\bf  r}'; \kappa)$
can be computed by applying
 translational invariance, i.e.,
$K_{{D}}({\bf r} ,{\bf r}';\kappa)
=
{\mathcal K}_{{D}}({\bf R};\kappa)$,
with ${\bf  R}={\bf  r} - {\bf  r}'$.
Its Fourier transform
$\widetilde{\mathcal K}_{{D}}({\bf  q};\kappa)
= -(q^{2}+\kappa^{2})^{-1}$
leads to an  integral expression
\begin{eqnarray}
{\mathcal K}_{{D}}({\bf  R};\kappa)
& = & -
\int \frac{ d^{{D}} q }{
(2 \pi )^{{D}} }
\frac{ e^{ i {\bf  q} \cdot {\bf  R} }}{q^{2}+\kappa^{2}
}
 \nonumber  \\
& = & -
 (2\pi)^{-{D}/2}
R^{-({D}/2 - 1)}
\int^{\infty}_{0}
\frac{q^{{D}/2}
J_{{D}/2 - 1}
(q R)}{q^{2}+\kappa^{2} } \, dq
\;  ,
\label{eq:DD_GF_mod}
\end{eqnarray}
in which Eq.~(\ref{eq:Bochner}) was used.
Equation~(\ref{eq:DD_GF_mod})  can be explicitly evaluated
in terms of the modified Bessel  
function of the second kind $K_{\nu} ( \kappa R)$, of order $\nu=D/2 - 1$,
i.e.~\cite{gra:80},
\begin{equation}
{\mathcal K}_{{D}}
({\bf  R};\kappa)
 =
- \frac{1}{2 \pi}
\left( \frac{\kappa}{2\pi R}
\right)^{\nu}
K_{\nu}(\kappa R)
\; ,
\label{eq:DD_GF2_mod}
\end{equation}
where the dimensional variable
 $\nu=D/2 - 1$ (cf.\ Eq.~(\ref{eq:nu_def}))
has been explicitly introduced.

Likewise,
for the ordinary Helmholtz equation,
\begin{equation}
\left[
 \nabla_{{\bf  r},{D}}^{2} +
k^{2}
  \right]
G_{{D}} ({\bf  r},{\bf  r}';k)
=
\delta^{({D})}
({\bf  r} - {\bf  r}')
\;
\label{eq:Helmholtz}
\end{equation}
in infinite space, translational invariance implies that
$G_{{D}} ({\bf  r}, {\bf  r}'; k)=
{\mathcal G}_{{D}}({\bf  R};k)$,
with ${\bf  R}={\bf  r} - {\bf  r}'$.
However, its Fourier transform
$\widetilde{\mathcal G}_{{D}}({\bf  q};k)
= (k^{2}-q^{2})^{-1}$
leads to an ill-defined  integral expression that needs to be
evaluated by a prescription defining the boundary conditions at 
infinity;
for outgoing $(+)$ and incoming $(-)$ boundary conditions,
\begin{eqnarray}
{\mathcal G}^{(\pm)}_{{D}}({\bf  R};k)
& = &
\int \frac{ d^{{D}} q }{ (2 \pi )^{{D}} }
\frac{ e^{ i {\bf  q} \cdot {\bf  R} }}{k^{2}-q^{2}\pm i\delta}
\nonumber \\
& = &
 (2\pi)^{-{D}/2} R^{-({D}/2 - 1)}
\int^{\infty}_{0}
\frac{q^{{D}/2}
J_{{D}/2 - 1}
(q R)}{k^{2}-q^{2} \pm i \delta}dq
\;  ,
\label{eq:DD_GF}
\end{eqnarray}
where Eq.~(\ref{eq:Bochner}) was used and $\delta=0^{+}$.
Equation~(\ref{eq:DD_GF})
can be explicitly evaluated in terms of Hankel functions
of order $\nu=D/2 - 1$;
in fact, it is easy to see that
Eq.~(\ref{eq:Helmholtz}) can be obtained from (\ref{eq:Helmholtz_mod})
with the replacement $\kappa= \mp i k$,
so that
\begin{equation}
{\mathcal G}^{(\pm)}_{{D}}({\bf  R};k)
=
{\mathcal K}_{{D}}({\bf  R};\kappa= \mp ik)
\;  ,
\label{eq:DD_GF_vs_DD_GF_mod}
\end{equation}
and the choice of signs amounts to the choice of boundary conditions
at infinity or the $i\delta $ prescription.
From the identity~\cite{abr:72}
\begin{equation}
K_{\nu}(\mp iz) =\pm \frac{\pi i}{2} e^{\pm i \pi \nu/2}
H^{(1,2)}_{\nu}(z)
\; ,
\end{equation}
Eq.~(\ref{eq:DD_GF_vs_DD_GF_mod}) is converted into
\begin{equation}
{\mathcal G}^{(\pm)}_{{D}}
({\bf  R};k)
 =
\mp \frac{i}{4}
\left( \frac{k}{2\pi R} \right)^{\nu}
H^{(1,2)}_{\nu}(kR)
\; .
\label{eq:DD_GF2}
\end{equation}
Equations~(\ref{eq:DD_GF2_mod}) and (\ref{eq:DD_GF2})
are well known~\cite{som:64}
and reduce to the familiar results in
one, two, and three dimensions~\cite{mor:53d}.

\section{SCATTERING IN $D$ DIMENSIONS}
\label{sec:D-dim_scattering}

Just as in the standard 3-D
 scattering formalism, the $D$-dimensional
time-independent
operator Schr\"{o}dinger equation
\begin{equation}
\left( H_{0} - E \right) |\Psi>
= -  V  |\Psi>
\;
\label{eq:OSE}
\end{equation}
is equivalent to
a Lippmann-Schwinger equation~\cite{sak:85}
\begin{equation}
|\Psi^{(+)}>
= |\chi >
+ \left( E-H_{0} + i \delta \right)^{-1}
 V |\Psi^{(+)}>
\;
\label{eq:OLS}
\end{equation}
in which the state vector
 $|\Psi^{(+)}> $ is explicitly resolved
into  an incident wave $|\chi>$ (solution of the free-particle case)
and a second term
that represents the outgoing scattered wave
(with an appropriate boundary condition
summarized by the $i\delta=i 0^{+}$ prescription).
In what follows, we will assume that $|\chi>=|\chi_{{\bf k}}>$
represents a $D$-dimensional
plane wave $e^{i {\bf k} \cdot {\bf r} }$.
Equation~(\ref{eq:OLS}) can be converted into the integral
form
\begin{equation}
\Psi^{(+)} ({\bf r})
= e^{i {\bf k} \cdot {\bf r}}
+
 \int d^{{D}}
r' G^{(+)}_{{D}} ({\bf  r}, {\bf  r}';k)
     V({\bf  r}') \Psi^{(+)} ({\bf  r}')
\; ,
\label{eq:LS}
\end{equation}
by the introduction of one of the
Green's functions  computed in Appendix~\ref{sec:D-dim}, namely,
$ G^{(+)}_{{D}} ({\bf  r}, {\bf  r}';k)=
<{\bf  r}
|\left( E-H_{0} + i \delta \right)^{-1}  |
 {\bf  r}'>
$, which is a  solution to Eq.~(\ref{eq:Helmholtz}),
and explicitly
given by Eq.~(\ref{eq:DD_GF2}).

For the scattering problem, one is interested in the
asymptotic form of the Green's
function, which follows from~\cite{gra:80}
\begin{equation}
H^{(1)}_{\nu} (z)
\stackrel{\scriptscriptstyle
(z \rightarrow \infty)}{\sim}
\sqrt\frac{2}{\pi z} \,  e^{i(z-\nu \pi/2 - \pi/4)}
\;  ,
\end{equation}
whence
\begin{equation}
G^{(+)}_{{D}}
({\bf  r}, {\bf r}';k)
=
{\mathcal G}^{(+)}_{{D}}
({\bf  R};k)
 \stackrel{\scriptscriptstyle
(r \rightarrow \infty)}{\sim}
- \frac{1}{4\pi}
\left( \frac{k}{2 \pi} \right)^{({D}-3)/2}
\, e^{i\gamma_{{D}}}
\,
\frac{e^{ikr}
}{r^{({D} -1)/2}} \, e^{-i {\bf  k}' \cdot {\bf  r}'}
\; ,
\label{eq:DD_GF_asymptotic}
\end{equation}
where ${\bf  k}'= k \, {\bf  r}/r$ and
 \begin{equation}
\gamma_{{D}}= \left( 3-D \right) \frac{\pi}{4}
\; .
\end{equation}
From Eqs.~(\ref{eq:LS})
 and (\ref{eq:DD_GF_asymptotic}),
in the position representation,
\begin{equation}
\Psi^{(+)} ({\bf  r})
\stackrel{\scriptscriptstyle (r \rightarrow \infty)}{\sim}
e^{i {\bf k} \cdot {\bf r} }
+
f^{(D)}_{k}
( \Omega^{({D})}   )
\, e^{i\gamma_{{D}}}
\,
\frac{e^{ikr}}{r^{({D} -1)/2}}
\;  ,
\end{equation}
where the scattering amplitude
\begin{equation}
f^{(D)}_{k}(\Omega^{({D})}) =
- \frac{1}{4 \pi} \left( \frac{k}{2 \pi}
\right)^{({D}-3)/2}
 \int d^{{D}} r' e^{-i {\bf  k}' \cdot {\bf  r}'}
     V({\bf  r}') \Psi^{(+)} ({\bf  r}')
\;
\label{eq:DD_LS_asymptotic}
\end{equation}
leads to the usual expression for the differential scattering
cross section,
\begin{equation}
\frac{d \sigma^{(D)}
(k, \Omega^{(D)} )}{d
\Omega_{D}}
= |f^{(D)}_{k}  ( \Omega^{({D})}   )
|^{2}
\;  .
\label{eq:diff_cross_section}
\end{equation}

\bigskip

\begin{center}
{\small\bf ACKNOWLEDGEMENTS}
\end{center}

This research was supported in part by
CONICET and ANPCyT, Argentina
(L.N.E., H.F.,
and C.A.G.C.) and by
the University of San Francisco Faculty Development Fund
(H.E.C.).
H.E.C. acknowledges
instructive discussions with
Professor Carlos R. Ord\'{o}\~{n}ez
and the generous hospitality of the University of Houston
during the final stage of preparation of this article.


\begin{references}
\bibitem{cam:00b_remark}
The entries in Table~\ref{tab:dimcontinuation}
for the two-dimensional delta-function potential
are computed in Section~\ref{sec:delta_DR}.
We defer the computational details for the inverse square potential
to Ref.~\cite{cam:00b}.
\end{references}
\end{document}